# Gaussian Quadrature Inference for Multicarrier Continuous-Variable Quantum Key Distribution

Laszlo Gyongyosi

[1] Quantum Technologies Laboratory, Department of Telecommunications
*Budapest University of Technology and Economics*
2 Magyar tudosok krt, Budapest, *H*-1117, Hungary
[2] MTA-BME Information Systems Research Group
*Hungarian Academy of Sciences*
7 Nador st., Budapest, *H*-1051, Hungary

gyongyosi@hit.bme.hu

**Abstract**

We propose the Gaussian quadrature inference (GQI) method for multicarrier continuous-variable quantum key distribution (CVQKD). A multicarrier CVQKD protocol utilizes Gaussian subcarrier quantum continuous variables (CV) for information transmission. The GQI framework provides a minimal error estimate of the quadratures of the CV quantum states from the discrete, measured noisy subcarrier variables. GQI utilizes the fundamentals of regularization theory and statistical information processing. We characterize GQI for multicarrier CVQKD, and define a method for the statistical modeling and processing of noisy Gaussian subcarrier quadratures. We demonstrate the results through the adaptive multicarrier quadrature division (AMQD) scheme. We define direct GQI (DGQI), and prove that it achieves a theoretical minimal magnitude error. We introduce the terms statistical secret key rate and statistical private classical information, which quantities are derived purely by the statistical functions of GQI. We prove the secret key rate formulas for a multiple access multicarrier CVQKD via the AMQD-MQA (multiuser quadrature allocation) scheme. The GQI and DGQI frameworks can be established in an arbitrary CVQKD protocol and measurement setting, and are implementable by standard low-complexity statistical functions, which is particularly convenient for an experimental CVQKD scenario.

# 1 Introduction

The continuous-variable quantum key distribution (CVQKD) protocols provide a plausible solution to practically realize an unconditional secure communication over standard, currently established telecommunication networks [10–22]. An important attribute of CVQKD is that, in contrast to DV (discrete variable) QKD, it does not require single-photon sources and detectors, and can be implemented by standard optical telecommunication devices [1], [9–26], [30–37]. In a CVQKD system, the information is carried by a continuous-variable quantum state that is defined in the phase space via the position and momentum quadratures. In practice, the CV quantum states have a Gaussian random distribution, because a Gaussian modulation is a considerable and well-established technique in an experimental scenario. The quantum channel between the sender (Alice) and receiver (Bob) is also provably Gaussian, because the presence of an eavesdropper (Eve) adds a white Gaussian noise into the transmission [19-21].

The CVQKD protocols have several attractive properties, however, the relevant performance attributes, such as secret key rates and transmission distances, still require significant improvements. For this purpose, the multicarrier CVQKD has been recently introduced through the adaptive quadrature division modulation (AMQD) scheme [2]. Precisely, the multicarrier CVQKD injects several additional degrees of freedom onto the transmission, which is not available for a standard, single-carrier CVQKD setting. In particular, these extra benefits and resources allow the realization of higher secret key rates and higher amount of tolerable losses with unconditional security. These innovations opened a door to the establishment of several new phenomena for CVQKD which are unrealizable in standard CVQKD, such as singular layer transmission [4], enhanced security thresholds [5], multidimensional manifold extraction [6], characterization of the subcarrier domain [7], adaptive quadrature detection and sub-channel estimation techniques [8], and an extensive utilization of distribution statistics and random matrix formalism [9]. The benefits of multicarrier CVQKD has also been proposed for multiple access multicarrier CVQKD via the AMQD-MQA (multiuser quadrature allocation) [3].

Statistical information processing is a statistical theory to extract information from signals. Statistical information processing has a wide application range from information theory to communication systems to physics [25-28]. In traditional communications, statistical information processing is a useful tool to characterize input signals from noisy observations, to achieve noise reduction, signal classification and compression, etc. The field of multirate statistical information processing deals with statistical information extraction from data that are characterized at non-equal sampling rates [28]. Statistical inference is an application of probability theory to propose generalized, plausible conclusions about a non-observable process. Particularly, the calculations are deductions and the conclusions are inferences based on observations. The inference rules utilize empirical methods to generate plausible results that are the subject of a desired solution. Statistical inference methods are rooted in the fundamentals of regularization theory. The practical aim of regularization theory is to provide a sequence of well-posed solutions that converges to an expected answer. The maximum entropy principle is a general method of statistical inference. It allows us to infer a probability distribution given certain constraints on the probability distribution itself.

In this work, we define a statistical information processing model of multicarrier CVQKD. We study the statistical attributes of the transmission of Gaussian CV quantum states, and define the method of *Gaussian Quadrature Inference* (GQI) for multicarrier CVQKD. The aim of GQI is to provide a statistical estimation of the input Gaussian subcarrier quadratures from the observed noisy Gaussian subcarriers, conveyed via the Gaussian sub-channels. Specifically, the GQI method is processing on the discrete noisy subcarrier quadrature components to recover the input Gaussian CV state in a continuous regime. The GQI method developed for multicarrier CVQKD utilizes the theory of mathematical statistics and the fundamentals of statistical information processing. We define *direct GQI* (DGQI), which is a flexible version of GQI. We prove that the GQI method achieves a theoretically minimized magnitude error, allowing one to determine the continuous variable Gaussian quadratures from the discrete variables with a vanishing error probability. We define the terms *statistical secret key rate* and *statistical private information*, and using the statistical functions of GQI we prove the corresponding formulas. We demonstrate the proofs through the AMQD-MQA (multiuser quadrature allocation) multiple access multicarrier CVQKD scheme.

The GQI and DGQI frameworks offer a minimal magnitude error, and are implementable by standard low-complexity functions. The integrated statistical functions are flexible, allowing it to be established in an arbitrary CVQKD protocol setting (one-way, two-way CVQKD) and measurement apparatuses (homodyne, heterodyne measurement), which is particularly convenient in an experimental CVQKD setting.

This paper is organized as follows. In Section 2, some preliminary findings are summarized. Section 3 discusses the Gaussian quadrature inference method for multiple access multicarrier CVQKD via the framework of AMQD-MQA. Section 4 provides the proof of the achievable statistical secret key rate in a GQI multicarrier CVQKD scenario. Finally, Section 5 concludes the results. Supplementary information is included in the Appendix.

## 2 Preliminaries

In Section 2, the notations and basic terms are summarized. For further information, see the detailed descriptions of [2–8].

### 2.1 Multicarrier CVQKD

The following description assumes a single user, and the use of $n$ Gaussian sub-channels $\mathcal{N}_i$ for the transmission of the subcarriers, from which only $l$ sub-channels will carry valuable information.

In the single-carrier modulation scheme, the $j$-th input single-carrier state $\left|\varphi_j\right\rangle = \left|x_j + \mathrm{i}p_j\right\rangle$ is a Gaussian state in the phase space $\mathcal{S}$, with i.i.d. Gaussian random position and momentum quad-

ratures $x_j \in \mathbb{N}\left(0, \sigma_{\omega_0}^2\right)$, $p_j \in \mathbb{N}\left(0, \sigma_{\omega_0}^2\right)$, where $\sigma_{\omega_0}^2$ is the modulation variance of the quadratures. In the multicarrier scenario, the information is carried by Gaussian subcarrier CVs, $\left|\phi_i\right\rangle = \left|x_i + \mathrm{i}p_i\right\rangle$, $x_i \in \mathbb{N}\left(0, \sigma_\omega^2\right)$, $p_i \in \mathbb{N}\left(0, \sigma_\omega^2\right)$, where $\sigma_\omega^2$ is the modulation variance of the subcarrier quadratures, which are transmitted through a noisy Gaussian sub-channel $\mathcal{N}_i$. Precisely, each $\mathcal{N}_i$ Gaussian sub-channel is dedicated for the transmission of one Gaussian subcarrier CV from the $n$ subcarrier CVs. (*Note*: index $i$ refers to a subcarrier CV, index $j$ to a single-carrier CV, respectively.)

The single-carrier CV state $\left|\varphi_j\right\rangle$ in the phase space $\mathcal{S}$ can be modeled as a zero-mean, circular symmetric complex Gaussian random variable $z_j \in \mathcal{CN}\left(0, \sigma_{\omega_{z_j}}^2\right)$, with a variance

$$\sigma_{\omega_{z_j}}^2 = \mathbb{E}\left[\left|z_j\right|^2\right] = 2\sigma_{\omega_0}^2, \tag{1}$$

and with i.i.d. real and imaginary zero-mean Gaussian random components

$$\mathrm{Re}\left(z_j\right) \in \mathbb{N}\left(0, \sigma_{\omega_0}^2\right), \ \mathrm{Im}\left(z_j\right) \in \mathbb{N}\left(0, \sigma_{\omega_0}^2\right). \tag{2}$$

In the multicarrier CVQKD scenario, let $n$ be the number of Alice's input single-carrier Gaussian states. Precisely, the $n$ input coherent states are modeled by an $n$-dimensional, zero-mean, circular symmetric complex random Gaussian vector

$$\mathbf{z} = \mathbf{x} + \mathrm{i}\mathbf{p} = \left(z_0, \ldots, z_{n-1}\right)^T \in \mathcal{CN}\left(0, \mathbf{K_z}\right), \tag{3}$$

where each $z_j$ is a zero-mean, circular symmetric complex Gaussian random variable

$$z_j \in \mathcal{CN}\left(0, \sigma_{\omega_{z_j}}^2\right), \ z_j = x_j + \mathrm{i}p_j. \tag{4}$$

In the first step of AMQD, Alice applies the inverse FFT (fast Fourier transform) operation to vector $\mathbf{z}$ (see (3)), which results in an $n$-dimensional zero-mean, circular symmetric complex Gaussian random vector $\mathbf{d}$, $\mathbf{d} \in \mathcal{CN}\left(0, \mathbf{K_d}\right)$, $\mathbf{d} = \left(d_0, \ldots, d_{n-1}\right)^T$, precisely as

$$\mathbf{d} = F^{-1}\left(\mathbf{z}\right) = e^{\frac{\mathbf{d}^T \mathbf{AA}^T \mathbf{d}}{2}} = e^{\frac{\sigma_{\omega_0}^2 \left(d_0^2 + \ldots + d_{n-1}^2\right)}{2}}, \tag{5}$$

where

$$d_i = x_{d_i} + \mathrm{i}p_{d_i}, \ d_i \in \mathcal{CN}\left(0, \sigma_{d_i}^2\right), \tag{6}$$

where $\sigma_{\omega_{d_i}}^2 = \mathbb{E}\left[\left|d_i\right|^2\right] = 2\sigma_\omega^2$, thus the position and momentum quadratures of $\left|\phi_i\right\rangle$ are i.i.d. Gaussian random variables with a constant variance $\sigma_\omega^2$ for all $\mathcal{N}_i, i = 0, \ldots, l-1$ sub-channels:

$$\mathrm{Re}\left(d_i\right) = x_{d_i} \in \mathbb{N}\left(0, \sigma_\omega^2\right), \ \mathrm{Im}\left(d_i\right) = p_{d_i} \in \mathbb{N}\left(0, \sigma_\omega^2\right), \tag{7}$$

where $\mathbf{K_d} = \mathbb{E}\left[\mathbf{dd}^\dagger\right]$, $\mathbb{E}\left[\mathbf{d}\right] = \mathbb{E}\left[e^{\mathrm{i}\gamma}\mathbf{d}\right] = \mathbb{E}e^{\mathrm{i}\gamma}\left[\mathbf{d}\right]$, and $\mathbb{E}\left[\mathbf{dd}^T\right] = \mathbb{E}\left[e^{\mathrm{i}\gamma}\mathbf{d}\left(e^{\mathrm{i}\gamma}\mathbf{d}\right)^T\right] = \mathbb{E}e^{\mathrm{i}2\gamma}\left[\mathbf{dd}^T\right]$

for any $\gamma \in [0, 2\pi]$.

The $\mathbf{T}(\mathcal{N})$ transmittance vector of $\mathcal{N}$ in the multicarrier transmission is

$$\mathbf{T}(\mathcal{N}) = [T_0(\mathcal{N}_0),...,T_{n-1}(\mathcal{N}_{n-1})]^T \in \mathcal{C}^n, \qquad (8)$$

where

$$T_i(\mathcal{N}_i) = \mathrm{Re}(T_i(\mathcal{N}_i)) + \mathrm{i}\,\mathrm{Im}(T_i(\mathcal{N}_i)) \in \mathcal{C}, \qquad (9)$$

is a complex variable, which quantifies the position and momentum quadrature transmission (i.e., gain) of the $i$-th Gaussian sub-channel $\mathcal{N}_i$, in the phase space $\mathcal{S}$, with real and imaginary parts

$$0 \leq \mathrm{Re}\,T_i(\mathcal{N}_i) \leq 1/\sqrt{2}, \text{ and } 0 \leq \mathrm{Im}\,T_i(\mathcal{N}_i) \leq 1/\sqrt{2}. \qquad (10)$$

Particularly, the $T_i(\mathcal{N}_i)$ variable has the squared magnitude of

$$|T_i(\mathcal{N}_i)|^2 = \mathrm{Re}\,T_i(\mathcal{N}_i)^2 + \mathrm{Im}\,T_i(\mathcal{N}_i)^2 \in \mathbb{R}, \qquad (11)$$

where

$$\mathrm{Re}\,T_i(\mathcal{N}_i) = \mathrm{Im}\,T_i(\mathcal{N}_i). \qquad (12)$$

The Fourier-transformed transmittance of the $i$-th sub-channel $\mathcal{N}_i$ (resulted from CVQFT operation at Bob) is denoted by

$$|F(T_i(\mathcal{N}_i))|^2. \qquad (13)$$

The $n$-dimensional zero-mean, circular symmetric complex Gaussian noise vector $\Delta \in \mathcal{CN}(0, \sigma_\Delta^2)_n$, of the quantum channel $\mathcal{N}$, is evaluated as

$$\Delta = (\Delta_0,...,\Delta_{n-1})^T \in \mathcal{CN}(0, \mathbf{K}_\Delta), \qquad (14)$$

where

$$\mathbf{K}_\Delta = \mathbb{E}[\Delta \Delta^\dagger], \qquad (15)$$

with independent, zero-mean Gaussian random components

$$\Delta_{x_i} \in \mathbb{N}(0, \sigma_{\mathcal{N}_i}^2), \text{ and } \Delta_{p_i} \in \mathbb{N}(0, \sigma_{\mathcal{N}_i}^2), \qquad (16)$$

with variance $\sigma_{\mathcal{N}_i}^2$, for each $\Delta_i$ of a Gaussian sub-channel $\mathcal{N}_i$, which identifies the Gaussian noise of the $i$-th sub-channel $\mathcal{N}_i$ on the quadrature components $x_i, p_i$ in the phase space $\mathcal{S}$. Thus $F(\Delta) \in \mathcal{CN}(0, \sigma_{\Delta_i}^2)$, where

$$\sigma_{\Delta_i}^2 = 2\sigma_{\mathcal{N}_i}^2. \qquad (17)$$

The CVQFT-transformed noise vector can be rewritten as

$$F(\Delta) = (F(\Delta_0),...,F(\Delta_{n-1}))^T, \qquad (18)$$

with independent components $F(\Delta_{x_i}) \in \mathbb{N}(0, \sigma_{\mathcal{N}_i}^2)$ and $F(\Delta_{p_i}) \in \mathbb{N}(0, \sigma_{\mathcal{N}_i}^2)$ on the quadra-tures, for each $F(\Delta_i)$. Precisely, it also defines an $n$-dimensional zero-mean, circular symmetric complex Gaussian random vector $F(\Delta) \in \mathcal{CN}(0, \mathbf{K}_{F(\Delta)})$ with a covariance matrix

$$\mathbf{K}_{F(\Delta)} = \mathbb{E}\left[F(\Delta)F(\Delta)^\dagger\right]. \tag{19}$$

The complex $A_j(\mathcal{N}_j) \in \mathbb{C}$ single-carrier channel coefficient is derived from the $l$ Gaussian sub-channel coefficients as

$$A_j(\mathcal{N}_j) = \tfrac{1}{l}\sum_{i=0}^{l-1} F(T_i(\mathcal{N}_i)). \tag{20}$$

The general model of AMQD is depicted in Fig. 1, for the details see [2].

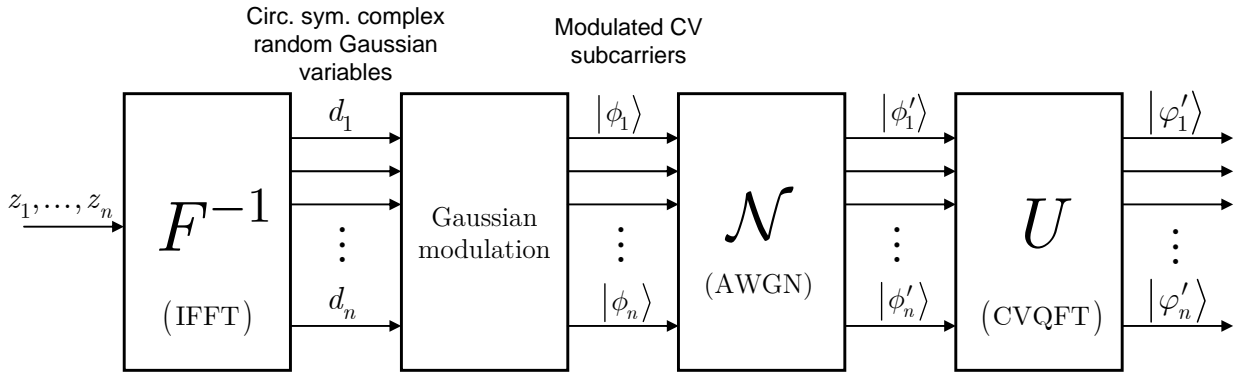

**Figure 1.** The AMQD modulation scheme [2]. Alice draws an $n$-dimensional, zero-mean, circular symmetric complex Gaussian random vector $\mathbf{z}$, which are then inverse Fourier-transformed by $F^{-1}$. The resulting vector $\mathbf{d}$ encodes the subcarrier quadratures for the Gaussian modulation. In the decoding, Bob applies the $U$ unitary CVQFT on the $n$ subcarriers to recover the noisy version of Alice's original variable as a continuous variable in the phase space (IFFT – inverse fast Fourier transform, AWGN – additive white Gaussian noise, CVQFT – inverse continuous-variable quantum Fourier transform).

### 2.1.1 Multiuser Quadrature Allocation (MQA) for Multicarrier CVQKD

In a MQA multiple access multicarrier CVQKD, a given user $U_k, k = 0,\ldots,K-1$, where $K$ is the number of total users, is characterized via $m$ subcarriers, formulating an $\mathcal{M}_{U_k}$ logical channel of $U_k$,

$$\mathcal{M}_{U_k} = \left[\mathcal{N}_{U_k,0},\ldots,\mathcal{N}_{U_k,m-1}\right]^T, \tag{21}$$

where $\mathcal{N}_{U_k,i}$ is the $i$-th sub-channel of $\mathcal{M}_{U_k}$.

For a detailed description of MQA for multicarrier CVQKD see [3], for the derivation of the security thresholds and secret key rate formulas, see [5].

The general model of AMQD-MQA is depicted in Fig. 2 [3].

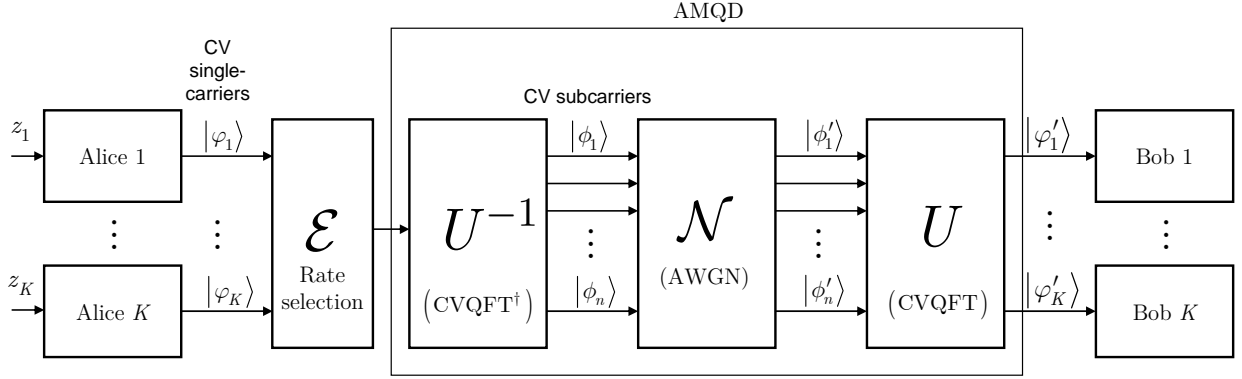

**Figure 2.** The AMQD-MQA multiple access scheme with multiple independent transmitters and multiple receivers [3]. The modulated Gaussian CV single-carriers are transformed by a unitary operation (inverse CVQFT) at the $\mathcal{E}$ encoder, which outputs the $n$ Gaussian subcarrier CVs for the transmission. The parties send the $|\varphi_k\rangle$ single-carrier Gaussian CVs with variance $\sigma_{\omega_{0,k}}^2$ to Alice. In the rate-selection phase, the encoder determines the transmit users. The data states of the transmit users are then fed into the CVQFT$^\dagger$ operation. The $|\phi_i\rangle$ Gaussian subcarrier CVs have a variance $\sigma_\omega^2$ per quadrature components. The Gaussian CVs are decoded by the CVQFT unitary operation. Each $|\varphi_k'\rangle$ is received by Bob $k$.

## 2.2 Statistical Information Processing

### 2.2.1 Basic Terms

#### 2.2.1.1 Wide-sense Stationary Processes

A WSS (wide-sense stationary) process $x(n)$ is a stochastic process [25-28], such that
$$\mathbb{E}\big[x^2(n)\big] < \infty, \forall n, \tag{22}$$
and
$$\mathbb{E}\big[x(n)\big] = C, \forall n, \tag{23}$$
where $C$ is a constant, and
$$\mathbb{E}\big[\big(x(n+k)-C\big)\big(x(n)-C\big)\big] = \mathbb{E}\big[\big(x(k)-C\big)\big(x(0)-C\big)\big], \forall n, k. \tag{24}$$

#### 2.2.1.2 Autocorrelation Function

Assuming that in (23),
$$C = \mathbb{E}\big[x(n)\big] = 0, \tag{25}$$
leads to the $\mathcal{A}_{x(n)}(\cdot)$ autocorrelation function (sequence) of $x(n)$ via (24) as
$$\mathcal{A}_{x(n)}(k) = \mathbb{E}\big[x(n+k)x(n)\big]. \tag{26}$$
It can be verified that
$$\mathcal{A}_{x(n)}(0) \geq 0, \tag{27}$$

$$\mathcal{A}_{x(n)}(-k) = \mathcal{A}_{x(n)}(k), \tag{28}$$

$$\left|\mathcal{A}_{x(n)}(k)\right| \leq \mathcal{A}_{x(n)}(0), \tag{29}$$

and $\mathcal{A}_{x(n)}(\cdot)$ is a non-negative definite sequence,

$$\sum_{i=1}^{n}\sum_{j=1}^{n} a_i a_j \mathcal{A}_{x(n)}(i-j) \geq 0, \forall n, \forall a_i, \tag{30}$$

where $a_i$ are real numbers, and

$$\det \mathbf{C}_{xx} \geq 0, \tag{31}$$

where $\mathbf{C}_{xx}$ is an $n \times n$ covariance matrix associated with $x(n)$, evaluated as

$$\left[\mathbf{C}_{xx}\right]_{ij} = \mathcal{A}_{x(n)}(i-j), i \geq 1, j \leq n. \tag{32}$$

#### 2.2.1.3  Entropy Rate of a Gaussian WSS

For a Gaussian WSS $x(n)$, the $H(x)$ entropy rate is as

$$H(x) = \tfrac{1}{2}\ln 2\pi + \tfrac{1}{2} + \tfrac{1}{4\pi}\int_{-\pi}^{\pi} \ln \mathcal{P}_x\left(e^{i\omega}\right) d\omega, \tag{33}$$

where $\mathcal{P}_x\left(e^{i\omega}\right)$ is the power spectrum of $x(n)$.

#### 2.2.1.4  Power Spectrum

The $\mathcal{P}_x\left(e^{i\omega}\right)$ is the power spectrum of $x(n)$ is expressed as

$$\mathcal{P}_x\left(e^{i\omega}\right) = \sum_{l=-\infty}^{+\infty} \mathcal{A}_{x(n)}(l) e^{-i\omega l}, \tag{34}$$

where $\omega \in [-\pi, \pi]$.

From (34) follows, that $\mathcal{P}_x\left(e^{i\omega}\right)$ is a measure of strength of the fluctuations of the Fourier components at a given $\omega$, allowing to write

$$\mathcal{P}_x\left(e^{i\omega}\right) = \mathcal{S}_x\left(e^{i\omega}\right), \tag{35}$$

where $\mathcal{S}_x\left(e^{i\omega}\right)$ is the spectral density of $x(n)$.

For $\mathcal{P}_x\left(e^{i\omega}\right)$,

$$f_R(\omega) = \mathcal{P}_x\left(e^{i\omega}\right), \tag{36}$$

where $f_R(\cdot)$ is a real function, and

$$\mathcal{P}_x\left(e^{-i\omega}\right) = \mathcal{P}_x\left(e^{i\omega}\right), \tag{37}$$

along with

$$\mathcal{P}_x\left(e^{i\omega}\right) \geq 0. \tag{38}$$

It also can be verified that for a Gaussian WSS $x(n)$,

$$H(x) = H(\mathcal{P}_x),\tag{39}$$

since $H(\cdot)$ is a functional of $\mathcal{P}_x(\cdot)$ [25-28], and

$$H(\mathcal{S}_x) = H(x) = H(\mathcal{P}_x).\tag{40}$$

#### 2.2.1.5 Regular Process

For an $x_r(n)$ regular process, $\mathcal{P}_x(e^{i\omega})$ is assumed to be a continuous function of $\omega$, such that

$$\int_{-\pi}^{\pi} \ln \mathcal{P}_{x_r}(e^{i\omega}) d\omega > -\infty.\tag{41}$$

### 2.2.2 Inference

Statistical inference is a tool of mathematical statistics to propose generalized, plausible conclusions from observations. The inference rules utilize empirical methods.

By theory, the aim of the maximum entropy principle is to infer an unknown function, $f(x)$, defined on a set $X$, if only a feasible set, $Y$, of such functions is available [28].

Let the $n$-tuple

$$(x_1,\ldots,x_n) \in X^n \tag{42}$$

drawn independently from a finite set $X$ of size $|X|$. Then, by the fundamentals of maximal entropy principle, the $f_E(x)$ empirical density function is

$$f_E(x) = \tfrac{1}{n}\bigl(|x_i = x|\bigr),\tag{43}$$

where $|\cdot|$ stands for the number of $x_i$ such that $x_i = x$.

The result in (43) has a probability of $p$, as

$$p = e^{-nD_X(f_E(x)\|q(x))-r_n(f_E(x))},\tag{44}$$

where $q(x)$ stands for the probability density at which $X$ is governed, and

$$0 \leq r_n(f_E(x)) \leq |X|\log n,\tag{45}$$

while $D_X(\cdot)$ is the classical relative entropy function, defined between probability density functions $p_1(\cdot)$ and $p_2(\cdot)$ on a countable set $\mathcal{C}$ as

$$D_X(p_1\|p_2) = \sum_{\mathcal{C}} p_1(X)\log_2 \tfrac{p_1(X)}{p_2(X)}.\tag{46}$$

For the $H(f_E(x))$ entropy function of $f_E(x)$,

$$N_n(f_E(x)) = e^{nH(f_E(x))-r_n(f_E(x))},\tag{47}$$

where $N_n(f_E(x))$ quantifies the number of $n$-tuples $(x_1,\ldots,x_n) \in X^n$ with a given $f_E(x)$.

### 2.2.3 Lebesgue Space and Functional Hilbert Space

The space of Lebesgue-measurable functions is defined as
$$\mathfrak{L}^p(a,b), \tag{48}$$
where $p < \infty$, and
$$f(x) \in \mathfrak{L}^p(a,b). \tag{49}$$
The $\mathfrak{L}^p$ norm with respect to $f(x)$ is expressed as
$$\|f\|_p = \tfrac{1}{b-a} \sqrt[p]{\int_a^b |f(x)|^p \, dx}, \tag{50}$$
while $\mathfrak{L}^\infty$ is evaluated as
$$\|f\|_\infty = \sup_{x \in (a,b)} |f(x)|. \tag{51}$$
The $\mathcal{H}_f(a,b)$ functional Hilbert space, $a = -\pi, b = \pi$, is defined as
$$\mathcal{H}_f(-\pi,\pi) = \mathfrak{L}^2(-\pi,\pi), \tag{52}$$
with a norm
$$\|f\|_2 = \tfrac{1}{2\pi} \sqrt{\int_{-\pi}^{\pi} |f(x)|^2 \, dx}. \tag{53}$$

## 3 Gaussian Quadrature Inference (GQI)

**Proposition 1** (Statistical modeling of Gaussian CV quantum states in a multicarrier CVQKD). *The Gaussian quadrature components of the CV quantum states statistically modeled as Gaussian WSS processes.*

The $x_{U_k,j}$ single-carrier Gaussian CV quadrature component is statistically modeled as an $x(n)$ Gaussian WSS process. The $\mathcal{M}_{U_k}(x_{U_k,j}) = x'_{U_k,j}$ noisy single-carrier quadrature components, and the $\mathcal{N}_{U_k,i}(x_{U_k,i}) = x'_{U_k,i}$, $i = 0, \ldots, m-1$ noisy Gaussian subcarrier CVs are also equivalent to Gaussian WSS processes, where $\mathcal{N}_{U_k,i}$ is the $i$-th Gaussian sub-channel component of the logical channel $\mathcal{M}_{U_k}$ (see (21)).

The optimal (least-squares) estimate of $x_{U_k,j}$ is $f_E(x_{U_k,j})$, where $f_E(\cdot)$ is a linear operator, with coefficients depending on $\mathcal{P}_{x_{U_k,j}}$. In a multicarrier CVQKD setting, the term $\mathcal{P}_{x_{U_k,j}}$ with respect to quadrature component $x_{U_k,j}$ can be approached as
$$\mathcal{P}_{x_{U_k,j}} = E\left(U^{-1}(x_{U_k,j})\right), \tag{54}$$
where $E(\cdot)$ stands for the estimator function, $U^{-1}$ is the inverse CVQFT operation.

The aim of GQI is to provide the continuous $E\left(U^{-1}\left(x_{U_k,j}\right)\right)$ from the discrete subcarrier components $x'_{U_k,i}$.

A variable $x'_{U_k,i}$ refers to a Gaussian quadrature component (real variable, position or momentum quadrature) of the $i$-th noisy subcarrier, resulting from a measurement operator $M$ (homodyne or heterodyne measurement, respectively).

## 3.1 GQI for Multiple Access Multicarrier CVQKD

**Theorem 1** (Gaussian Quadrature Inference for multicarrier CVQKD). *The $m$ $x'_{U_k,i}$, $i = 0,\dots,m-1$, noisy subcarrier CVs of $U_k$, $k = 0,\dots,K-1$, yield the $E\left(U^{-1}\left(x_{U_k,j}\right)\right)$ estimate of $U^{-1}\left(x_{U_k,j}\right)$, where $x_{U_k,j}$ is the quadrature component of $\varphi_{U_k,j}$, $\varphi_{U_k,j}$ is the $j$-th input CV of $U_k$, $\varphi_{U_k,j} = x_{U_k,j} + \mathrm{i}p_{U_k,j}$, $\{x_{U_k,j}, p_{U_k,j}\}$ are Gaussian random quadratures, as*

$$E\left(U^{-1}\left(x_{U_k,j}\right)\right) = \left|1\bigg/-\sum_{i=0}^{m-1}\left|\mathcal{T}_i\left(e^{\mathrm{i}\theta_{\varphi_{U_k,j}}}\right)\right|^2\tilde{\lambda}_i\right|,$$

*where function $\mathcal{T}_i\left(e^{\mathrm{i}\theta_{\varphi_{U_k,j}}}\right)$ evaluates $\mathcal{T}_{U_k,i}\left(\mathcal{N}_{U_k,i}\right)$ of $\mathcal{N}_{U_k,i}$, $\mathcal{N}_{U_k,i}$ is the $i$-th sub-channel of $U_k$, while $\tilde{\lambda}_i$ are optimal Lagrange multipliers.*

*Proof.*
Let

$$\varphi_{U_k,j} = x_{U_k,j} + \mathrm{i}p_{U_k,j}, \tag{55}$$

be the single-carrier Gaussian CV of $U_k$, $k = 0,\dots,K-1$, $\varphi_{U_k,j} \in \mathbb{N}\left(0, \mathbb{E}\left\|\varphi_{U_k,j}\right\|^2 = 2\sigma^2_{\omega_0}\right)$, where

$$x_{U_k,j} \in \mathbb{N}\left(0,\sigma^2_{\omega_0}\right),\ p_{U_k,j} \in \mathbb{N}\left(0,\sigma^2_{\omega_0}\right) \tag{56}$$

are Gaussian random quadratures, $\sigma^2_{\omega_0}$ is the single-carrier modulation variance [2], and let the $m$ subcarrier CVs of $U_k$ be referred via

$$\vec{\phi}_{U_k} = \left[\phi_{U_k,0}\dots\phi_{U_k,m-1}\right]^T, \tag{57}$$

where

$$\phi_{U_k,i} = x_{U_k,i} + \mathrm{i}p_{U_k,i}, \tag{58}$$

$\phi_{U_k,i} \in \mathbb{N}\left(0, \mathbb{E}\left\|\phi_{U_k,i}\right\|^2 = 2\sigma^2_{\omega_i}\right)$, while

$$x_{U_k,i} \in \mathbb{N}\left(0,\sigma^2_{\omega_i}\right),\ p_{U_k,i} \in \mathbb{N}\left(0,\sigma^2_{\omega_i}\right) \tag{59}$$

are the subcarrier quadratures, $\sigma^2_{\omega_i}$ is the quadrature modulation variance (chosen to be constant $\sigma^2_{\omega_i} = \sigma^2_\omega$ for $\forall i$, see [2]), while $\mathcal{M}_{U_k}$ is the logical channel (a set of $m$ sub-channels) of $U_k$, see (21).

The output of $\mathcal{N}_{U_k,i}$ is $\phi'_{U_k,i} \in \mathbb{N}\left(0, 2\sigma^2_{\omega_i} + \sigma^2_{\mathcal{N}_{U_k,i}}\right)$, where $\sigma^2_{\mathcal{N}_{U_k,i}}$ is the noise variance of $\mathcal{N}_{U_k,i}$, and

$$\overrightarrow{\phi'}_{U_k} = \left[\phi'_{U_k,0} \cdots \phi'_{U_k,m-1}\right]^T, \tag{60}$$

where

$$\phi'_{U_k,i} = x'_{U_k,i} + \mathrm{i} p'_{U_k,i} \tag{61}$$

and $x'_{U_k,i}, p'_{U_k,i}$ are noisy Gaussian random quadratures,

$$x'_{U_k,i} \in \mathbb{N}\left(0, \sigma^2_\omega + \sigma^2_{\Delta_{U_k,i}}\right),\ p'_{U_k,i} \in \mathbb{N}\left(0, \sigma^2_\omega + \sigma^2_{\Delta_{U_k,i}}\right), \tag{62}$$

where $\sigma^2_{\Delta_{U_k,i}}$ is the quadrature-level noise variance of $\mathcal{N}_{U_k,i}$, thus $\sigma^2_{\mathcal{N}_{U_k,i}} = 2\sigma^2_{\Delta_{U_k,i}}$.

Note the proof is demonstrated for a single quadrature component of (55), thus allowing us to use $x_{U_k,j}$ of $\varphi_{U_k,j}$ in the remaining parts.

Let $n$ be the number of single-carriers, $n \to \infty$, and let

$$\theta_{\varphi_{U_k,j}} = \pi/\Omega, \tag{63}$$

where

$$\Omega = \sigma^2_{\omega_0} / \sigma^2_\omega, \tag{64}$$

and where $\sigma^2_{\omega_0}$, $\sigma^2_\omega$ are the single-carrier and multicarrier modulation variances, respectively.

Statistically, in a multicarrier CVQKD setting, the following relation can be written between a single-carrier $x_{U_k,j}$ and subcarrier $x_{U_k,i}$ Gaussian quadrature component (assuming $\Omega = 1$):

$$x_{U_k,i} = \sum_{j=-\infty}^{\infty} x_{U_k,j} e^{\mathrm{i}j\theta_{\varphi_{U_k,j}}}, \tag{65}$$

and

$$x_{U_k,j} = \frac{1}{2\pi} \int_{-\pi}^{\pi} x_{U_k,i} e^{-\mathrm{i}t_j \theta_{\varphi_{U_k,j}}} d\theta_{\varphi_{U_k,j}}. \tag{66}$$

Specifically, for any $\sigma^2_\omega < \sigma^2_{\omega_0}$, it follows that $\Omega \neq 1$ and $\left|\theta_{\varphi_{U_k,j}}\right| < \pi$, therefore $x_{U_k,i}$ in (65) can be rewritten as

$$x_{U_k,i} = \frac{1}{\Omega} \sum_{j=-\infty}^{\infty} x_{U_k,j} e^{\mathrm{i}j\theta_{\varphi_{U_k,j}} \frac{1}{\Omega}}. \tag{67}$$

Note that in (65) it is assumed that the integral of (66) exists and is invertible, thus $x_{U_k,j}$ is either square-integrable or absolutely integrable [28].

The $x'_{U_k,i}$ noisy version of (65) is available for Bob via a corresponding $M$ measurement operator (e.g., homodyne or heterodyne measurement) performed on the noisy $\phi'_i$ CV state, as

$$x'_{U_k,i} = M(\phi'_i) = M(\mathcal{N}_{U_k,i}(\phi_i)). \tag{68}$$

In particular, the $\mathcal{S}\left(e^{-i\theta_{\varphi_{U_k,j}}}\right)$ spectral density of $x_{U_k,j}$ can be defined via the $\left|x'_{U_k,i}\right|^2$ expectation value of $x'_{U_k,i}$, as

$$\mathcal{S}\left(e^{-i\theta_{\varphi_{U_k,j}}}\right) = \mathbb{E}\left(\left|x'_{U_k,i}\right|^2\right), \tag{69}$$

which is a statistical measure of the strength of the fluctuations of the subcarrier components [2], [28].

Precisely, it can be verified that (69) is analogous to the power spectrum $\mathcal{P}\left(e^{-i\theta_{\varphi_{U_k,j}}}\right)$ of $x_{U_k,j}$,

$$\mathcal{S}\left(e^{-i\theta_{\varphi_{U_k,j}}}\right) = \mathcal{P}\left(e^{-i\theta_{\varphi_{U_k,j}}}\right), \tag{70}$$

where $\mathcal{P}\left(e^{-i\theta_{\varphi_{U_k,j}}}\right) \geq 0$ is a real function of $\theta_{\varphi_{U_k,j}}$,

$$\mathcal{P}\left(e^{-i\theta_{\varphi_{U_k,j}}}\right) = \sum_{g=-\infty}^{\infty} \mathcal{A}_{x_{U_k,j}}(g) e^{-i\theta_{\varphi_{U_k,j}} g}, \tag{71}$$

such that

$$\mathcal{P}\left(e^{-i\theta_{\varphi_{U_k,j}}}\right) = \mathcal{P}\left(e^{i\theta_{\varphi_{U_k,j}}}\right), \tag{72}$$

where $\mathcal{A}_{x_{U_k,j}}(\cdot)$ is the autocorrelation function (autocorrelation sequence [25-28]) of $x_{U_k,j}$, expressed as

$$\mathcal{A}_{x_{U_k,j}}(g) = \mathbb{E}\left(x_{U_k,j+g} x_{U_k,j}\right). \tag{73}$$

Without loss of generality, (69) and (71), allow us to write

$$\mathcal{S}\left(e^{i\theta_{\varphi_{U_k,j}}}\right) = \mathbb{E}\left(\left|x'_{U_k,i}\right|^2\right). \tag{74}$$

Using (74), the estimation of $U^{-1}\left(x_{U_k,j}\right)$, where $U^{-1}(\cdot)$ is the inverse CVQFT unitary operation, is expressed as

$$E\left(U^{-1}\left(x_{U_k,j}\right)\right) = \mathcal{S}\left(e^{i\theta_{\varphi_{U_k,j}}}\right) = \mathcal{P}\left(e^{i\theta_{\varphi_{U_k,j}}}\right), \tag{75}$$

which, by using (71) can be further evaluated as

$$E\left(U^{-1}\left(x_{U_k,j}\right)\right) = \sum_{g=-\infty}^{\infty} \mathcal{A}_{x_{U_k,j}}(g) e^{-i\theta_{\varphi_{U_k,j}} g} \qquad (76)$$
$$= \mathbb{E}\left(\left|x'_{U_k,i}\right|^2\right).$$

In particular, $E\left(U^{-1}\left(x_{U_k,j}\right)\right)$ allows us to uniquely specify $\mathcal{A}_{x'_{U_k,i}}(g)$ of a noisy subcarrier quadrature $x'_{U_k,i}$ as follows.

For a noisy subcarrier quadrature $x'_{U_k,i}$ of the $i$-th subcarrier CV $\phi'_{U_k,i}$ of $U_k$,
$$\mathcal{A}_{x'_{U_k,i}}(g) = \mathcal{A}_{x_{U_k,j}}(\Omega g), \qquad (77)$$
where $\Omega$ is defined in (64), while $\mathcal{A}_{x_{U_k,j}}(g)$ of $x_{U_k,j}$ is as

$$\begin{aligned}\mathcal{A}_{x_{U_k,j}}(g) &= \tfrac{1}{2\pi}\int_{-\pi}^{\pi} E\left(U^{-1}\left(x_{U_k,j}\right)\right) \mathcal{G}_i\left(e^{i\theta_{\varphi_{U_k,j}}}\right) e^{ig\theta_{\varphi_{U_k,j}}} d\theta_{\varphi_{U_k,j}} \\ &= \tfrac{1}{2\pi}\int_{-\pi}^{\pi} \mathcal{S}\left(e^{-i\theta_{\varphi_{U_k,j}}}\right) \mathcal{G}_i\left(e^{i\theta_{\varphi_{U_k,j}}}\right) e^{ig\theta_{\varphi_{U_k,j}}} d\theta_{\varphi_{U_k,j}} \\ &= \tfrac{1}{2\pi}\int_{-\pi}^{\pi} \mathcal{P}\left(e^{i\theta_{\varphi_{U_k,j}}}\right) \mathcal{G}_i\left(e^{i\theta_{\varphi_{U_k,j}}}\right) e^{ig\theta_{\varphi_{U_k,j}}} d\theta_{\varphi_{U_k,j}},\end{aligned} \qquad (78)$$

where $\mathcal{G}_i\left(e^{i\theta_{\varphi_{U_k,j}}}\right)$ is defined as
$$\mathcal{G}_i\left(e^{i\theta_{\varphi_{U_k,j}}}\right) = \mathcal{T}_i\left(e^{i\theta_{\varphi_{U_k,j}}}\right) \mathcal{T}_i\left(\tfrac{1}{e^{i\theta_{\varphi_{U_k,j}}}}\right), \qquad (79)$$
where
$$\mathcal{T}_i\left(e^{i\theta_{\varphi_{U_k,j}}}\right) = \begin{cases} T_i\left(\mathcal{N}_{U_k,i}\right), & \text{if } \left|\theta_{\varphi_{U_k,j}}\right| \leq \tfrac{\pi}{\Omega}, \\ 0, & \text{otherwise.} \end{cases} \qquad (80)$$

Note that (79) can also be determined via a pilot CV state-based channel estimation procedure; for details, see [8].

Without loss of generality, using (78), (77) can be rewritten as
$$\begin{aligned}\mathcal{A}_{x'_{U_k,i}}(g) &= \tfrac{1}{2\pi}\int_{-\pi}^{\pi} E\left(U^{-1}\left(x_{U_k,j}\right)\right) \mathcal{G}_i\left(e^{i\theta_{\varphi_{U_k,j}}}\right) e^{i\Omega g\theta_{\varphi_{U_k,j}}} d\theta_{\varphi_{U_k,j}} \\ &= \tfrac{1}{2\pi}\int_{-\pi}^{\pi} \mathcal{S}\left(e^{-i\theta_{\varphi_{U_k,j}}}\right) \mathcal{G}_i\left(e^{i\theta_{\varphi_{U_k,j}}}\right) e^{i\Omega g\theta_{\varphi_{U_k,j}}} d\theta_{\varphi_{U_k,j}} \\ &= \tfrac{1}{2\pi}\int_{-\pi}^{\pi} \mathcal{P}\left(e^{i\theta_{\varphi_{U_k,j}}}\right) \mathcal{G}_i\left(e^{i\theta_{\varphi_{U_k,j}}}\right) e^{i\Omega g\theta_{\varphi_{U_k,j}}} d\theta_{\varphi_{U_k,j}}.\end{aligned} \qquad (81)$$

According to the fundamentals of the maximum entropy principle,
$$\widetilde{E}\left(U^{-1}\left(x_{U_k,j}\right)\right) = \arg\max H\left(x_{U_k,j}\right), \qquad (82)$$
since $H(\cdot)$ is a functional of $\mathcal{P}_{x_{U_k,j}}(\cdot)$ (see (39)), subject to

$$E\left(U^{-1}\left(x_{U_k,j}\right)\right) \in \wp, \tag{83}$$

where set $\wp$ is defined as

$$\wp = \{\tfrac{1}{2\pi}\int_{-\pi}^{\pi}\mathcal{P}\left(e^{i\theta_{\varphi_{U_k,j}}}\right)\mathcal{G}_i\left(e^{i\theta_{\varphi_{U_k,j}}}\right)e^{i\Omega\theta_{\varphi_{U_k,j}}q}d\theta_{\varphi_{U_k,j}}, \forall x'_{U_k,i} \in X_{\phi'_{U_k,i}}, q = 0,\ldots,L-1,$$
$$\cup \mathcal{P}\left(e^{i\theta_{\varphi_{U_k,j}}}\right) \in \mathfrak{L}^1(-\pi,\pi), \tag{84}$$
$$\cup \mathcal{P}\left(e^{i\theta_{\varphi_{U_k,j}}}\right) \geq 0\},$$

where $\mathfrak{L}^1(\cdot)$ is a Lebesgue space, while set $X_{\phi'_{U_k,i}}$ is defined as the set of the autocorrelation functions of the $m$ subcarriers of $U_k$,

$$X_{\phi'_{U_k,i}} = \left\{\mathcal{A}_{x'_{U_k,0}},\ldots,\mathcal{A}_{x'_{U_k,m-1}}\right\}, \tag{85}$$

where set $X_{\phi'_{U_k,i}}$ is admissible, if $\wp \neq 0$, and $H\left(x_{U_k,j}\right)$ is the entropy rate [28] of the Gaussian quadrature component $x_{U_k,j}$ of $\varphi_{U_k,j}$, evaluated as

$$H\left(x_{U_k,j}\right) = \tfrac{1}{2}\ln 2\pi + \tfrac{1}{2} + \tfrac{1}{4\pi}\int_{-\pi}^{\pi}\ln\mathcal{P}\left(e^{-i\theta_{\varphi_{U_k,j}}}\right)d\theta_{\varphi_{U_k,j}}, \tag{86}$$

which is a concave functional, by theory [28].

Further it is assumed that (85) is available for Bob from the $x'_{U_k,i}$ measured subcarrier quadratures; see (66), along with $\mathcal{G}_i(x)$.

In particular, it can be verified that (87) can be rewritten as

$$\widetilde{E}\left(U^{-1}\left(x_{U_k,j}\right)\right) = \frac{1}{\sum_{i=0}^{m-1}\mathcal{G}_i\left(e^{i\theta_{\varphi_{U_k,j}}}\right)\mathcal{F}_i\left(e^{i\Omega\theta_{\varphi_{U_k,j}}}\right)}, \tag{88}$$

where $\mathcal{F}_i(x)$ is a transfer function

$$\mathcal{F}_i(x) = \sum_{q=-(L-1)}^{L-1} 2\lambda_{iq}x^{-q}, \tag{89}$$

where $\lambda_{iq}$ are the Lagrange multipliers, $i = 0,\ldots,m-1, q = 0,\ldots,L-1$, and

$$\lambda_{iq} = \lambda_{i(-q)}. \tag{90}$$

Note that in a CVQKD setting,

$$L = 1 \text{ and } q = 0, \tag{91}$$

since only one $M$ measurement operation is performed on a given subcarrier CV state $\phi'_{U_k,i}$ that yields a given variable $x'_{U_k,i}$ per each subcarrier CV state.

Set (84) can be reformulated with respect to $x_{U_k,j}$ as

$$\wp(q=0) = \{ \tfrac{1}{2\pi} \int_{-\pi}^{\pi} \mathcal{P}\!\left(e^{i\theta_{\varphi_{U_k,j}}}\right) \mathcal{G}_i\!\left(e^{i\theta_{\varphi_{U_k,j}}}\right) d\theta_{\varphi_{U_k,j}}, \forall x'_{U_k,i} \in \mathrm{X}_{\phi'_{U_k,i}},$$
$$\cup\, \mathcal{P}\!\left(e^{i\theta_{\varphi_{U_k,j}}}\right) \in \mathcal{L}^1(-\pi,\pi), \qquad (92)$$
$$\cup\, \mathcal{P}\!\left(e^{i\theta_{\varphi_{U_k,j}}}\right) \geq 0 \},$$

and

$$\mathcal{F}_i(x) = 2\lambda_i. \qquad (93).$$

Precisely, the $\lambda_i$ Lagrangian coefficient in (93) is determined via $\mathcal{A}_{x'_{U_k,i}}$ as

$$\mathcal{A}_{x'_{U_k,i}} = \frac{1}{2\pi} \int_{-\pi}^{\pi} \frac{\mathcal{G}_i\!\left(e^{i\theta_{\varphi_{U_k,j}}}\right)}{\sum_{u=0}^{m-1} \mathcal{G}_u\!\left(e^{i\theta_{\varphi_{U_k,j}}}\right) \mathcal{F}_u\!\left(e^{i\Omega\theta_{\varphi_{U_k,j}}}\right)} d\theta_{\varphi_{U_k,j}}. \qquad (94)$$

Since $\mathcal{F}_i(x)$ is a real function of $\theta_{\varphi_{U_k,j}}$, (89) at $q=0$ can be expressed as

$$\mathcal{F}_i(x) = \lambda_i. \qquad (95)$$

Using (79), for any $\left|\theta_{\varphi_{U_k,j}}\right| \leq \pi$, $\widetilde{E}\!\left(U^{-1}(x_{U_k,j})\right)$ from (88) can also be rewritten as

$$\widetilde{E}\!\left(U^{-1}(x_{U_k,j})\right) = \frac{1}{\sum_{i=0}^{m-1} \left|\mathcal{T}_i\!\left(e^{i\theta_{\varphi_{U_k,j}}}\right)\right|^2 \mathcal{F}_i\!\left(e^{i\Omega\theta_{\varphi_{U_k,j}}}\right)}, \qquad (88)$$

while $\mathcal{A}_{x'_{U_k,i}}$ from (94) is as

$$\mathcal{A}_{x'_{U_k,i}} = \frac{1}{2\pi} \int_{-\pi}^{\pi} \frac{\left|\mathcal{T}_i\!\left(e^{i\theta_{\varphi_{U_k,j}}}\right)\right|^2}{\sum_{u=0}^{m-1} \left|\mathcal{T}_u\!\left(e^{i\theta_{\varphi_{U_k,j}}}\right)\right|^2 \mathcal{F}_u\!\left(e^{i\Omega\theta_{\varphi_{U_k,j}}}\right)} d\theta_{\varphi_{U_k,j}}. \qquad (96)$$

If

$$E\!\left(U^{-1}(x_{U_k,j})\right) \notin \wp, \qquad (97)$$

then $\mathcal{A}_{x'_{U_k,i}}$ cannot be determined via (96), which requires to define a different constraint, $\Theta$, to find $\lambda_i$, such that

$$\Theta = \min \sum_{i=0}^{m-1} \sum_{q=0}^{L-1} \left( \frac{1}{2\pi} \int_{-\pi}^{\pi} \frac{\left|\mathcal{T}_i\!\left(e^{i\theta_{\varphi_{U_k,j}}}\right)\right|^2 e^{i\Omega\theta_{\varphi_{U_k,j}}q}}{\sum_{u=0}^{m-1} \left|\mathcal{T}_u\!\left(e^{i\theta_{\varphi_{U_k,j}}}\right)\right|^2 \mathcal{F}_u\!\left(e^{i\Omega\theta_{\varphi_{U_k,j}}}\right)} d\theta_{\varphi_{U_k,j}} - \mathcal{A}_{x'_{U_k,i}} \right)^2$$
$$= \min \sum_{i=0}^{m-1} \left( \frac{1}{2\pi} \int_{-\pi}^{\pi} \frac{\left|\mathcal{T}_i\!\left(e^{i\theta_{\varphi_{U_k,j}}}\right)\right|^2}{\sum_{u=0}^{m-1} \left|\mathcal{T}_u\!\left(e^{i\theta_{\varphi_{U_k,j}}}\right)\right|^2 \mathcal{F}_u\!\left(e^{i\Omega\theta_{\varphi_{U_k,j}}}\right)} d\theta_{\varphi_{U_k,j}} - \mathcal{A}_{x'_{U_k,i}} \right)^2, \qquad (98)$$

which is, in fact, a least-square approximation, that in contrast to (96) always exists [25-28]. Specifically, in function of the $\Gamma$ Lagrangian set

$$\Gamma = \left[\lambda_0, \ldots, \lambda_{(m-1)}\right]^T, \qquad (99)$$

(98) can be expressed as [28]
$$\Theta(\Gamma) = \sum_{i=0}^{m-1} \left(\omega_i(\Gamma) - \mathcal{A}_{x'_{U_k,i}}\right)^2, \tag{100}$$

where
$$\omega_i(\Gamma) = \frac{1}{2\pi} \int_{-\pi}^{\pi} \left| \frac{\left|\mathcal{T}_i\left(e^{i\theta_{\varphi_{U_k,j}}}\right)\right|^2 e^{i\Omega\theta_{\varphi_{U_k,j}} q}}{\sum_{u=0}^{m-1} \left|\mathcal{T}_u\left(e^{i\theta_{\varphi_{U_k,j}}}\right)\right|^2 \lambda_u \cos\left(\Omega\theta_{\varphi_{U_k,j}}\right)} \right| d\theta_{\varphi_{U_k,j}}$$
$$= \frac{1}{2\pi} \int_{-\pi}^{\pi} \left| \frac{\left|\mathcal{T}_i\left(e^{i\theta_{\varphi_{U_k,j}}}\right)\right|^2}{\sum_{u=0}^{m-1} \left|\mathcal{T}_u\left(e^{i\theta_{\varphi_{U_k,j}}}\right)\right|^2 \lambda_u \cos\left(\Omega\theta_{\varphi_{U_k,j}}\right)} \right| d\theta_{\varphi_{U_k,j}}. \tag{101}$$

Then, defining $\tilde{\omega}_i(\Gamma)$ as
$$\tilde{\omega}_i(\Gamma) = \arg\min \omega_i(\Gamma), \tag{102}$$

provides the $\tilde{\lambda}_i$ optimal Lagrangian coefficients such that (98) is satisfied.

Finally, using (88) and (93) leads to
$$E\left(U^{-1}\left(x_{U_k,j}\right)\right) = \left| \frac{1}{\sum_{i=0}^{m-1} \left|\mathcal{T}_i\left(e^{i\theta_{\varphi_{U_k,j}}}\right)\right|^2 \sum_{q=0}^{L-1} \tilde{\lambda}_{iq} \cos\left(\Omega\theta_{\varphi_{U_k,j}}\right)} \right|$$
$$= \left| \frac{1}{\sum_{i=0}^{m-1} \left|\mathcal{T}_i\left(e^{i\theta_{\varphi_{U_k,j}}}\right)\right|^2 \tilde{\lambda}_i \cos\left(\Omega\theta_{\varphi_{U_k,j}}\right)} \right| \tag{103}$$
$$= \left| \frac{1}{-\sum_{i=0}^{m-1} \left|\mathcal{T}_i\left(e^{i\theta_{\varphi_{U_k,j}}}\right)\right|^2 \tilde{\lambda}_i} \right|.$$

The result obtained in (103) is unique, since (86) is a concave functional [25-28].

To verify it, let $\mathcal{C}$ be a corresponding constraint set which contains $E^{(1)}\left(U^{-1}\left(x_{U_k,j}\right)\right)$ and $E^{(2)}\left(U^{-1}\left(x_{U_k,j}\right)\right)$, from which
$$E^{(\alpha)}\left(U^{-1}\left(x_{U_k,j}\right)\right) = \alpha E^{(1)}\left(U^{-1}\left(x_{U_k,j}\right)\right) + (1-\alpha) E^{(2)}\left(U^{-1}\left(x_{U_k,j}\right)\right) \tag{104}$$

can be defined.

In particular, for $E^{(\alpha)}\left(U^{-1}\left(x_{U_k,j}\right)\right)$ one can conclude that at $0 \leq \alpha \leq 1$,
$$E^{(\alpha)}\left(U^{-1}\left(x_{U_k,j}\right)\right) \in \mathcal{C}. \tag{105}$$

Then there exists an $f(\cdot)$, a corresponding function such that
$$f(\alpha) = f\left(E^{(\alpha)}\left(U^{-1}\left(x_{U_k,j}\right)\right)\right)$$
$$= \frac{1}{2\pi} \int_{-\pi}^{\pi} \ln\left(\alpha E^{(1)}\left(U^{-1}\left(x_{U_k,j}\right)\right) + (1-\alpha) E^{(2)}\left(U^{-1}\left(x_{U_k,j}\right)\right)\right) d\theta_{\varphi_{U_k,j}}, \tag{106}$$

and

$$\frac{\partial^2 f(\alpha)}{\partial \alpha^2} = \frac{1}{2\pi} \int_{-\pi}^{\pi} \frac{-\left(E^{(1)}\left(U^{-1}\left(x_{U_k,j}\right)\right) - E^{(2)}\left(U^{-1}\left(x_{U_k,j}\right)\right)\right)^2}{E^{(\alpha)}\left(U^{-1}\left(x_{U_k,j}\right)\right)^2} d\theta_{\varphi_{U_k,j}} < 0, \tag{107}$$

which shows that $\mathcal{C}$ is, indeed, a convex set [25-28]. Precisely, (86) is strictly concave, thus (103) has a maximum over the corresponding constraint set, and is always unique.

The solution (103) can be represented in the $\mathcal{H}_f(-\pi, \pi) = \mathcal{L}^2(-\pi, \pi)$ functional Hilbert space. Let us assume that $E\left(U^{-1}\left(x_{U_k,j}\right)\right) \in \mathcal{H}_f(A, B)$; thus the optimization problem can be reformulated as

$$\Pi = \min \int_A^B \delta E\left(U^{-1}\left(x_{U_k,j}\right)\right) d\theta_{\varphi_{U_k,j}}, \tag{108}$$

where $\delta(\cdot)$ is the corresponding entropy measure function, subject to

$$\Psi_i = \int_A^B f_i\left(\theta_{\varphi_{U_k,j}}\right) E\left(U^{-1}\left(x_{U_k,j}\right)\right) d\theta_{\varphi_{U_k,j}}, \tag{109}$$

where $f_i(\cdot)$ is a known function, $\Psi_i$ stands for the measured subcarrier CV quadrature, i.e., without loss of generality

$$\Psi_i = x'_{U_k,i}. \tag{110}$$

Exploiting some fundamental theory leads to the fact that an optimal entropy measure function in the $\mathcal{H}_f$ statistical space is the $H_B$ Burg entropy [28], evaluated via $\delta(\cdot)$ as

$$\delta(x) = H_B(x) = -\log x. \tag{111}$$

Then let

$$\ell = \left[c_0, c_1, c_2\right]^T \in \mathbb{R}^3, \tag{112}$$

and

$$\mathfrak{T} = \left\{\ell \,\middle|\, \langle \ell, r \rangle = 3, \ell \geq 0\right\}, \tag{113}$$

where $r = (1,1,1)^T$ is a reference vector in the $\mathbb{R}^3$ Euclidean space, while $\langle \ell, r \rangle = 3$ defines the $\Upsilon$ plane surface. In particular, $\mathfrak{T}$ represents those points on $\Upsilon$ that are confined to the first (positive) quadrant [28]. The $H_B$ Burg entropy can be defined to measure the information statistical distance for $\mathfrak{T}$ as

$$H_B = \sum_{i=0}^{2} \ln c_i, \tag{114}$$

which defines contours in $\mathfrak{T}$. Since $E\left(U^{-1}\left(x_{U_k,j}\right)\right)$ has the highest entropy among other possible solutions, its point $P \in \mathcal{H}_f(-\pi, \pi)$ will be the closest to the center $c$ of $\mathfrak{T}$ in $\mathcal{H}_f(-\pi, \pi)$, where $c$ has a maximal entropy. A non-optimal solution is denoted by $\hat{P} \in \mathcal{H}_f(-\pi, \pi)$.

The specification of $P$ in $\mathcal{H}_f(-\pi, \pi)$ is depicted in Fig. 3.

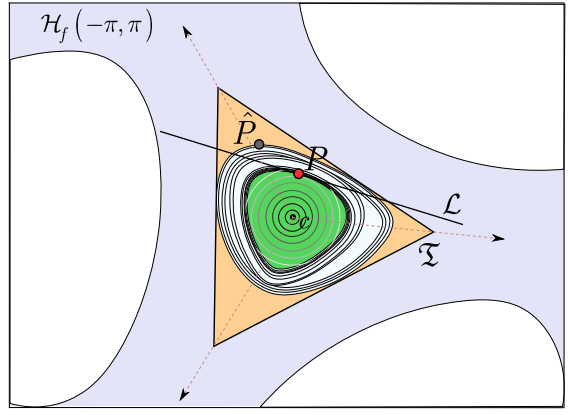

**Figure 3.** Representation of $E\left(U^{-1}\left(x_{U_k,j}\right)\right)$ via $P$ in $\mathcal{H}_f\left(-\pi,\pi\right)$. The line $\mathcal{L}$, defined by a linear constraint, determines $P$. The point $P$ has a maximal entropy among the set of possible solutions, $\hat{P}$. The distorted geometry is a consequence of generator function $H_B$.

Since (101) is non-tractable if $\mathcal{G}_i(x) \to 0$, in Theorem 2 we introduce a direct GQI method that evaluates $E\left(U^{-1}\left(x_{U_k,j}\right)\right)$ directly from the outputs of the $M$ operator.

∎

The GQI scheme for multicarrier CVQKD is illustrated in Fig. 4.

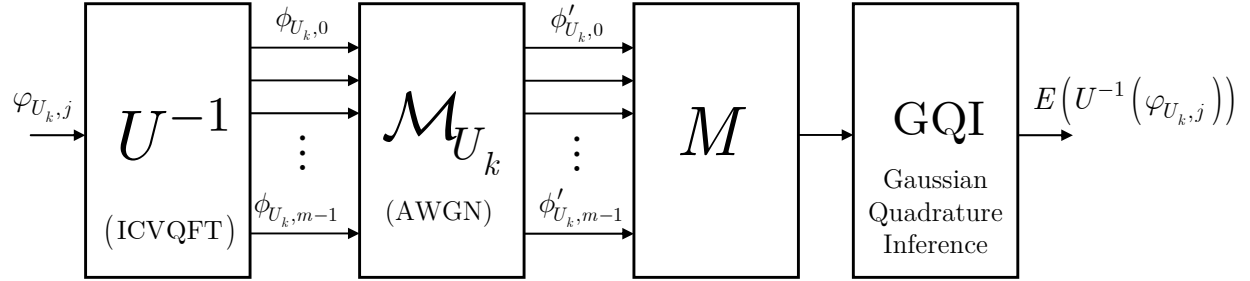

**Figure 4.** The Gaussian Quadrature Inference (GQI) for multicarrier CVQKD. User $U_k$ is equipped with a logical channel $\mathcal{M}_{U_k} = \left[\mathcal{N}_{U_k,0},...,\mathcal{N}_{U_k,m-1}\right]^T$, and $m$ subcarrier CVs. The input of $U_k$ is $\varphi_{U_k,j} = x_{U_k,j} + \mathrm{i}p_{U_k,j}$, which is transformed via the $U^{-1}$ ICVQFT operation. The $E\left(U^{-1}\left(\varphi_{U_k,j}\right)\right)$ estimate of the $j$-th input CV of $U_k$, $\varphi_{U_k,j} = x_{U_k,j} + \mathrm{i}p_{U_k,j}$ is yielded from the $m$ noisy Gaussian subcarrier CVs, $\phi'_{U_k,i} = x'_{U_k,i} + \mathrm{i}p'_{U_k,i}, i = 0,...,m-1$. (ICVQFT – inverse continuous-variable quantum Fourier transform). Note that the model above is analogous to the case where the input of $U_k$ is $z_{U_k,j} = x_{U_k,j} + \mathrm{i}p_{U_k,j}$ is a classical continuous variable which is transformed by the $F^{-1}$ IFFT operation (note that the quadratures are measured by the $M$ operation, e.g., via heterodyne or homodyne measurement, respectively).

## 3.2 Direct GQI (DGQI)

**Theorem 2** (Direct method of GQI). $E\left(U^{-1}\left(x_{U_k,j}\right)\right) = F^{-1}\left(x'_{U_k,j}\right) * \beta_i$, where $*$ is the linear convolution, function $\beta_i$ provides an $\varepsilon$ minimal magnitude error, $\varepsilon = \arg\min \varepsilon_{\max}$, where $\varepsilon_{\max} = \max\limits_{\forall x}\left||x_{U_k,j}|^2 - |x'_{U_k,j}|^2\right|$, $x_{U_k,j}$, $x'_{U_k,j}$ are the input, output single-carrier quadratures, $F^{-1}(\cdot)$ is the inverse FFT operation.

*Proof.*

The proof is proposed for a given quadrature component of the CV state, $x_{U_k,j}$.

Specifically, for the error analysis, we define quantity $\varepsilon_{\max}$ as

$$\begin{aligned}\varepsilon_{\max} &= \max_{\forall x}\left|\operatorname{Re}\mathbb{E}\left(|\varphi_{U_k,j}|^2\right) - \operatorname{Re}\mathbb{E}\left(|\varphi'_{U_k,j}|^2\right)\right| \\ &= \max_{\forall x}\left||x_{U_k,j}|^2 - |x'_{U_k,j}|^2\right|,\end{aligned} \quad (115)$$

which characterizes the maximal deviation of $|x'_{U_k,j}|^2$ from $|x_{U_k,j}|^2$.

Particularly, we define an abstracted offset range [29-30] of $\varepsilon_{\max}$, $\mho_{\sigma_\omega^2}$, as

$$\mho_{\sigma_\omega^2} : \left\{0 \leq \sigma_\omega^2 \leq 0.5\right\}, \quad (116)$$

Thus, (115) quantifies the maximal deviation from 0 dB at $0 \leq \sigma_\omega^2 \leq 0.5$.

Operator $A_i$ is defined precisely as

$$A_i = 2\sigma_\omega^2 \tfrac{i}{m}, \quad i = 0,\ldots,\tfrac{m}{2}, \quad (117)$$

where $\sigma_\omega^2$ is the subcarrier modulation variance and $m$ is the number of subcarriers of $U_k$.

Since in a multicarrier CVQKD setting, the $x'_{U_k,i}$ subcarriers are discretized via $M$, but $x_{U_k,i}$ and $x_{U_k,j}$ are continuous variables with a different modulation variance, the problem of discontinuity has to be resolved. The solution is as follows.

Let $x'_{U_k,i}$ be the $i$-th noisy subcarrier, $i = 0,\ldots,m-1$, and

$$F^{-1}\left(x'_{U_k,j}\right) = \sum_{i=0}^{m-1} x'_{U_k,z} e^{\frac{i2\pi ij}{m}}, j = 0,\ldots,n-1. \quad (118)$$

Then, let $\mathfrak{E}$ an estimator function [29-30], defined over the space of the $m$ subcarriers as

$$\mathfrak{E} = \begin{cases} i = 0 : \mathfrak{E}\left(A_0\right) = \tfrac{1}{m^2}\left|x'_{U_k,0}\right|^2 \\ i = 1,\ldots,\tfrac{m}{2} - 1 : \mathfrak{E}\left(A_i\right) = \tfrac{1}{m^2}\left(\left|x'_{U_k,i}\right|^2 + \left|x'_{U_k,m-i}\right|^2\right), \\ i = \tfrac{m}{2} : \mathfrak{E}\left(A_{\tfrac{m}{2}}\right) = \tfrac{1}{m^2}\left|x'_{U_k,\tfrac{m}{2}}\right|^2 \end{cases} \quad (119)$$

where $A_i$ is shown in (117).

The variables of the proof (single-carrier level with respect to quadrature $x_{U_k,j}$) are summarized in Fig. 5. The input single-carrier CV quadrature component is depicted by $x_{U_k,j}$, and the subcarrier CVs are referred via $U^{-1}(x_{U_k,j})$. The $M$ measurement operator (homodyne or heterodyne measurement) results in the discrete $F^{-1}(x'_{U_k,j})$, where $x'_{U_k,j}$ is the discretized, noisy version of the continuous variable $x_{U_k,j}$.

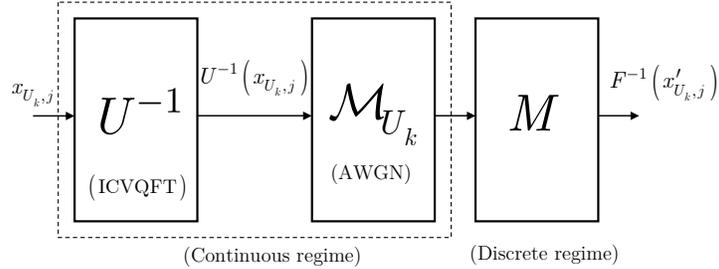

**Figure 5.** Single-carrier level variables with respect to quadrature $x_{U_k,j}$. The CV quantum states are defined in the continuous regime, while the $M$ measurement operator produces discrete variables.

By some fundamental theory, it can be verified that (119) provides a mean squared estimation of the CV state $F^{-1}(x_{U_k,j})$, as

$$\mathfrak{E}\left(F^{-1}(x_{U_k,j})\right) = \frac{1}{m}\sum_{i=0}^{m-1}\left|x'_{U_k,i}\right|^2$$
$$= \frac{1}{t}\int_0^t \left|x'_{U_k,i}(t)\right|^2 dt, \qquad (120)$$

thus $\mathfrak{E}(F(x_{U_k,j}))$ is an approximation of the continuous $F(x_{U_k,j})$ via $\left|x'_{U_k,i}\right|^2$ the subcarrier (squared) magnitudes. Since (119) is taken at discretized $A_i$, the quality of the approximation also depends on (117).

The discontinuity problem can be resolved by the definition of an appropriate function, $\mathfrak{f}$, which utilizes $A_i$ from (117), as follows.

It can be verified that the result in (120) is analogous to averaging $F(x_{U_k,j})$ over a narrow range function $\mathfrak{f}$, centered on a corresponding $A_i$, defined as

$$\mathfrak{f}(s) = \frac{1}{m^2}\left(\frac{\sin(\pi s)}{\sin\left(\frac{\pi s}{m}\right)}\right)^2 = \left|F^{-1}(\beta_i)\right|^2, \qquad (121)$$

where $s$ is an independent integer variable, and $\beta_i$ is a corresponding function, will be specified later.

To step further we apply the convolution theorem to evaluate $E(U^{-1}(x_{U_k,j}))$ from $F^{-1}(x'_{U_k,j})$ and $F^{-1}(\beta_i)$, as

$$F^{-1}\left(x'_{U_k,j} * \beta_i\right) = F^{-1}\left(x'_{U_k,j}\right) * F^{-1}\left(\beta_i\right)$$
$$= E\left(U^{-1}\left(x_{U_k,j}\right)\right), \tag{122}$$

where $*$ is the linear convolution operator [25-29].

In particular, it can be proven that (121) can be rewritten as [30]

$$\mathfrak{f}(s) = \frac{1}{\alpha}\left|\sum_{i=0}^{m-1} \beta_i e^{\frac{i2\pi is}{m}}\right|^2$$
$$= \frac{1}{\alpha}\left|\int_{-m/2}^{m/2} \cos\left(\frac{2\pi si}{m}\right)\beta\left(i - \frac{m}{2}\right)di\right|^2, \tag{123}$$

where

$$\alpha = m\sum_{i=0}^{m-1}\beta_i^2, \tag{124}$$

while function $\beta_i$ provides a range selection, defined later.

Without loss of generality, at a given $\beta_i$, $F^{-1}\left(x'_{U_k,j} * \beta_i\right)$ from (122) is as

$$F^{-1}\left(x'_{U_k,j} * \beta_i\right) = \sum_{i=0}^{m-1} \beta_i x'_{U_k,i} e^{\frac{i2\pi ij}{m}}, j = 0,\ldots,n-1, \tag{125}$$

and the results obtained in (120) can be expressed as

$$\mathfrak{E}(\beta_i) = \begin{cases} i = 0 : \mathfrak{E}(\beta_i, A_0) = \frac{1}{\alpha}\left|\sum_{z=0}^{m-1}\beta_z x'_{U_k,z}\right|^2 \\ i = 1,\ldots,\frac{m}{2}-1 : \mathfrak{E}(\beta_i, A_i) = \frac{1}{\alpha}\left(\left|\sum_{z=0}^{m-1}\beta_z x'_{U_k,z} e^{\frac{i2\pi zi}{m}}\right|^2 + \left|\sum_{z=0}^{m-1}\beta_z x'_{U_k,z} e^{\frac{i2\pi z(m-i)}{m}}\right|^2\right) \\ i = \frac{m}{2} : \mathfrak{E}(\beta_i, A_{\frac{m}{2}}) = \frac{1}{\alpha}\left|\sum_{z=0}^{m-1}\beta_z x'_{U_k,z} e^{i\pi z}\right|^2 \end{cases}. \tag{126}$$

As it can be concluded, a relevant point of the evolution of $F^{-1}\left(x'_{U_k,j} * \beta_i\right)$ is the value of $\beta_i$. Specifically, choosing

$$\beta_i = 1 \tag{127}$$

results in

$$\mathfrak{f}(s) = \frac{1}{\alpha}\left|\sum_{i=0}^{m-1} e^{\frac{i2\pi is}{m}}\right|^2$$
$$= \frac{1}{m^2}\left|\sum_{i=0}^{m-1} e^{\frac{i2\pi is}{m}}\right|^2, \tag{128}$$

which is a strictly sub-optimal solution, because $\beta_i = 1$ provides only a simple truncation of the continuous regime. It is provable that it can provide only a significantly distorted approximation of the continuous range of the input CV quantum state, leading to a maximization of $\varepsilon_{\max}$ in (115).

Since an experimental CVQKD protocol is operating in the low SNR regimes, the magnitude errors (see (115)) has a critical significance, and a subject of a strict minimization.

Precisely, the minimization of $\varepsilon_{\max}$ requires a careful selection of $\beta_i$. To reach the desired $\varepsilon$ error threshold,

$$\varepsilon = \arg\min \varepsilon_{\max}, \tag{129}$$

we characterize $\beta_{i,\varepsilon}$ specifically for a multicarrier CVQKD setting, as

$$\beta_{i,\varepsilon} = 1 + \sum_{y=1}^{P} C_y \cos(yQ_i), \tag{130}$$

where $C_0$ is arbitrarily set to unity [29], $P$ is the number of $C_y$ coefficients, while

$$Q_i = \tfrac{2\pi i}{m}, i = 0,\dots, m-1. \tag{131}$$

Using (130), leads to a closed form of $f_\varepsilon(s)$ via (123) as

$$\begin{aligned}
f_\varepsilon(s) &= \tfrac{1}{\alpha_\varepsilon} \left| \sum_{i=0}^{m-1} \left( 1 + \sum_{y=1}^{P} C_y \cos(yQ_i) \right) e^{\tfrac{\mathrm{i}2\pi i s}{m}} \right|^2 \\
&= \tfrac{1}{\alpha_\varepsilon} \left| \sum_{i=0}^{m-1} \left( e^{\tfrac{\mathrm{i}2\pi i s}{m}} + \sum_{y=1}^{P} C_y e^{\tfrac{\mathrm{i}2\pi i s}{m}} \cos\left(\tfrac{y2\pi i}{m}\right) \right) \right|^2,
\end{aligned} \tag{132}$$

where

$$\alpha_\varepsilon = m \sum_{z=0}^{m-1} \beta_{z,\varepsilon}^2. \tag{133}$$

Then, function $\mathfrak{E}_\varepsilon(\beta_i)$ in (126) can be rewritten as

$$\mathfrak{E}_\varepsilon(\beta_{i,\varepsilon}) = \begin{cases}
i = 0 : \mathfrak{E}_\varepsilon(\beta_{i,\varepsilon}, A_0) = \tfrac{1}{\alpha_\varepsilon} \left| \sum_{z=0}^{m-1} \beta_{i,\varepsilon} x'_{U_k,z} \right|^2 \\
i = 1,\dots,\tfrac{m}{2}-1 : \mathfrak{E}_\varepsilon(\beta_{i,\varepsilon}, A_i) = \tfrac{1}{\alpha_\varepsilon} \left( \left| \sum_{z=0}^{m-1} \beta_{i,\varepsilon} x'_{U_k,z} e^{\tfrac{\mathrm{i}2\pi z i}{m}} \right|^2 + \left| \sum_{z=0}^{m-1} \beta_{i,\varepsilon} x'_{U_k,z} e^{\tfrac{\mathrm{i}2\pi z(m-i)}{m}} \right|^2 \right), \\
i = \tfrac{m}{2} : \mathfrak{E}_\varepsilon(\beta_{i,\varepsilon}, A_{\tfrac{m}{2}}) = \tfrac{1}{\alpha_\varepsilon} \left| \sum_{z=0}^{m-1} \beta_{i,\varepsilon} x'_{U_k,z} e^{\mathrm{i}\pi z} \right|^2
\end{cases} \tag{134}$$

leading to a closed form of

$$\mathfrak{E}_\varepsilon\left(\beta_{i,\varepsilon}\right) = \begin{cases} i = 0 : \mathfrak{E}_\varepsilon\left(\beta_{i,\varepsilon}, \mathrm{A}_0\right) = \dfrac{1}{m\sum_{z=0}^{m-1}\beta_{z,\varepsilon}^2} \left|\sum_{z=0}^{m-1}\left(1 + \sum_{y=1}^{P} C_y \cos\left(yQ_i\right)\right) x'_{U_k,z}\right|^2 \\[2ex] i = 1,\ldots,\dfrac{m}{2} - 1 : \mathfrak{E}_\varepsilon\left(\beta_{i,\varepsilon}, \mathrm{A}_i\right) = \dfrac{1}{m\sum_{z=0}^{m-1}\beta_{z,\varepsilon}^2} \left( \left|\sum_{z=0}^{m-1}\left(1 + \sum_{y=1}^{P} C_y \cos\left(yQ_i\right)\right) x'_{U_k,z} e^{\frac{\mathrm{i}2\pi zi}{m}}\right|^2 \right. \\[2ex] \qquad\qquad \left. + \left|\sum_{z=0}^{m-1}\left(1 + \sum_{y=1}^{P} C_y \cos\left(yQ_i\right)\right) x'_{U_k,z} e^{\frac{\mathrm{i}2\pi z(m-i)}{m}}\right|^2 \right) \\[2ex] i = \dfrac{m}{2} : \mathfrak{E}_\varepsilon\left(\beta_{i,\varepsilon}, \mathrm{A}_{\frac{m}{2}}\right) = \dfrac{1}{m\sum_{z=0}^{m-1}\beta_{z,\varepsilon}^2} \left|\sum_{z=0}^{m-1}\left(1 + \sum_{y=1}^{P} C_y \cos\left(yQ_i\right)\right) x'_{U_k,z} e^{\mathrm{i}\pi z}\right|^2 \end{cases}.$$

(135)

Note, the optimal value of $\beta_{i,\varepsilon}$ is determined by nonlinear optimization methods through the iteration of coefficients $C_y$, $y = 1,\ldots,P$ (see (130)).

In particular, the conditions for the boundary continuity and differentiation [29] of $\beta_{i,\varepsilon}$ can be satisfied via

$$1 + \sum_{y=1}^{P} C_y = 0. \tag{136}$$

Defining an $\mathrm{N}(\cdot)$ normalization term for $\beta_i$ and $\beta_{i,\varepsilon}$,

$$\mathrm{N}(\beta_i) = \beta_i \frac{m}{\sum_{i=0}^{m-1}\beta_i}, \text{ and } \mathrm{N}(\beta_{i,\varepsilon}) = \beta_{i,\varepsilon} \frac{m}{\sum_{i=0}^{m-1}\beta_{i,\varepsilon}}, \tag{137}$$

the various $\beta_i$ functions become comparable.

The performance of DGQI can be exactly characterized via the magnitude error, $\varepsilon_{\max}$, at a given $\mathrm{N}(\beta_i)$.

The $\mathrm{N}(\beta_i)$ and $\mathrm{N}(\beta_{i,\varepsilon})$ normalized results for $\beta_i = 1$ and $\beta_{i,\varepsilon}$, where $\varepsilon = \arg\min \varepsilon_{\max}$, in function of $i/m$ are depicted in Fig. 6.

At $\beta_i = 1$, $\mathrm{N}(\beta_i) = 1$ for any $i/m$, while for $\beta_{i,\varepsilon}$, $\mathrm{N}(\beta_{i,\varepsilon})$ is maximized at $i/m = 0.5$.

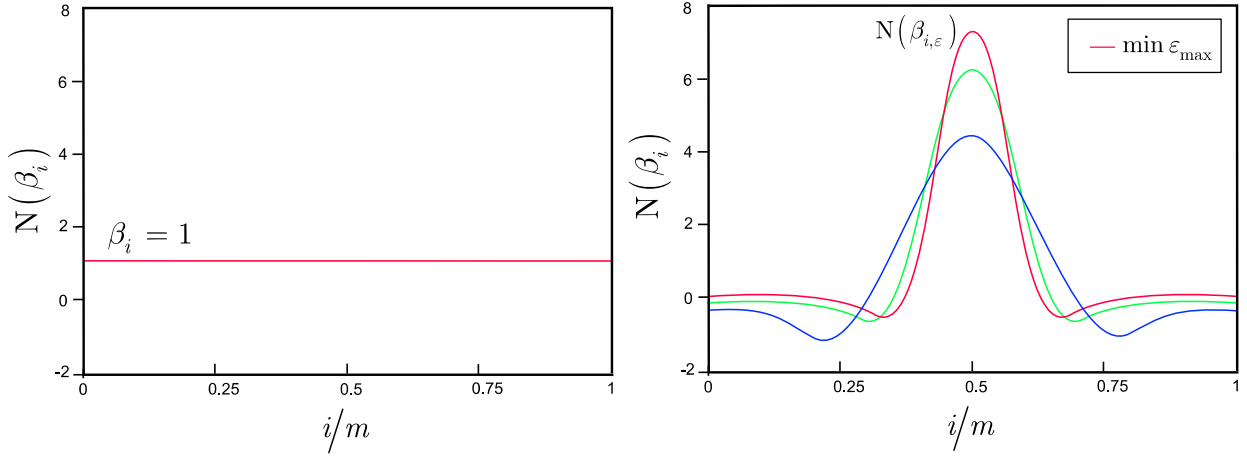

**Figure 6.** Evolution of $\beta_i$ of DGQI. The normalized $N(\beta_i)$ values for $\beta_i = 1$ and $\beta_{i,\varepsilon}$, $\varepsilon = \arg\min \varepsilon_{\max}$, in function of $i/m$. The $\varepsilon$ minimized magnitude error is obtained at $N(\beta_{i,\varepsilon})$.

The values of $\varepsilon_{\max}$ (dB) in function of $\mho_{\sigma_\omega^2}$ (scaled for the range $0 \leq \mho_{\sigma_\omega^2} \leq 0.5$) for $\beta_i = 1$ and $\beta_{i,\varepsilon}$ are depicted in Fig. 7.

The $\varepsilon_{\max}$ maximal magnitude error at $\beta_{i,\varepsilon}$ is almost negligible, $\varepsilon_{\max} \approx 0$, $\beta_{i,\varepsilon}$ is evaluated via the iteration of $C_y$, $y = 1,\ldots,P$, coefficients of (130). At a range truncation, $\beta_i = 1$ (see (127)), $\varepsilon_{\max}$ picks up an extremal value.

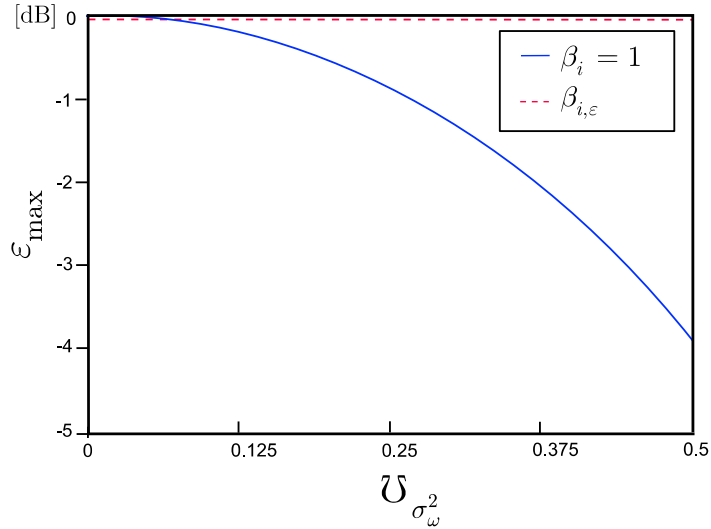

**Figure 7.** The $\varepsilon_{\max}$ maximal magnitude error (dB) of DGQI, in function of $\mho_{\sigma_\omega^2}$, for $\beta_i = 1$ and $\beta_{i,\varepsilon}$. At $\beta_{i,\varepsilon}$, $\varepsilon_{\max}$ is almost zero for an arbitrary $\mho_{\sigma_\omega^2}$.

∎

The direct GQI (DGQI) scheme is summarized in Fig. 8.

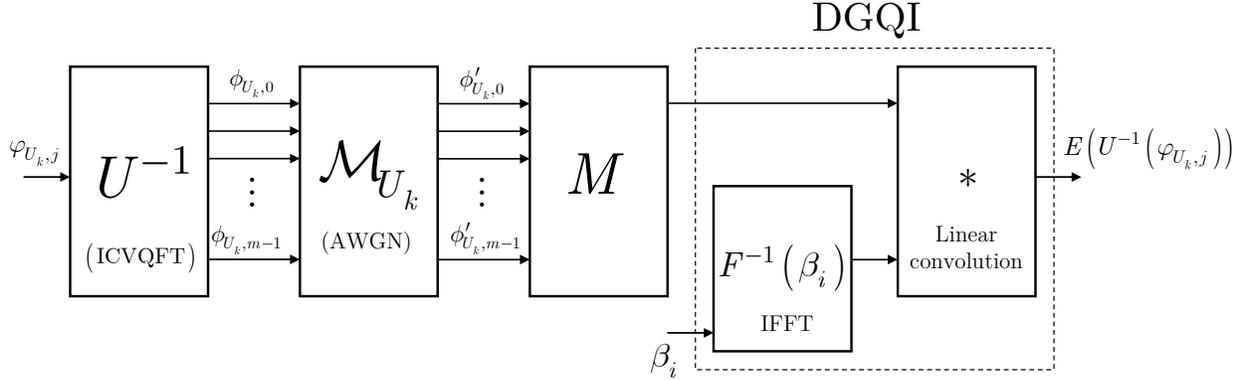

**Figure 8.** The DGQI method. Function $\beta_i$ is transformed by the $F^{-1}(\cdot)$ inverse FFT operation. The output of $M$ and $F^{-1}(\beta_i)$ are convolved, to obtain $E\left(U^{-1}\left(\varphi_{U_k,j}\right)\right)$.

# 4 Statistical Secret Key Rate

In a CVQKD setting, it can be proven that the amount of information leaked to an eavesdropper is theoretically minimized at a reverse reconciliation. We also propose the results of Theorem 3 for the case of reverse reconciliation. The results can be trivially extended for direct reconciliation.

**Theorem 3** (Statistical secret key rate). *Let $\hat{x}_{(j;Z),U_k}$, $\hat{x}'_{(j;Z),U_k}$, and $\hat{x}'_{(j;Z),E}$ be the optimal quadratures of Alice, Bob and Eve, obtained at $Z$ autocorrelation coefficients. The $S\left(\mathcal{M}_{U_k}\right)$ statistical secret key rate of $U_k$ in a multicarrier CVQKD at a reverse reconciliation is*

$$\begin{aligned}
S\left(\mathcal{M}_{U_k}\right) &\leq \lim_{n\to\infty} \frac{1}{n} P\left(\mathcal{M}_{U_k}\right) \\
&= \lim_{n\to\infty} \frac{1}{n} \max_{\forall p_i, \rho_i} \left(\chi_{AB} - \chi_{BE}\right) \\
&= \lim_{n\to\infty} \frac{1}{n} \left(\min_{\forall \sigma} \max_{\forall \rho} D\left(\rho_k^{AB} \| \sigma^{AB}\right) - \min_{\forall \sigma} \max_{\forall \rho} D\left(\rho_k^{BE} \| \sigma^{BE}\right)\right) \\
&= \lim_{n\to\infty} \frac{1}{n} \lim_{Z\to\infty} \max_{\forall x} \left(D_{AB}\left(\hat{x}'_{(j;Z),U_k} \| \hat{x}_{(j;Z),U_k}\right) - D_{BE}\left(\hat{x}'_{(j;Z),E} \| \hat{x}'_{(j;Z),U_k}\right)\right) \\
&= \lim_{n\to\infty} \frac{1}{n} \lim_{Z\to\infty} \max_{\forall x} \frac{1}{4\pi} \left(\int_{-\pi}^{\pi} \left(\mathcal{P}_{\hat{x}'_{(j;Z),U_k}}\left(e^{i\theta_{\varphi_{U_k,j}}}\right) - \ln \mathcal{P}_{\hat{x}'_{(j;Z),U_k}}\left(e^{i\theta_{\varphi_{U_k,j}}}\right) - 1\right) d\theta_{\varphi_{U_k,j}} \right. \\
&\quad \left. - \left(\int_{-\pi}^{\pi} \left(\mathcal{P}_{\hat{x}'_{(j;Z),E}}\left(e^{i\theta_{\varphi_{U_k,j}}}\right) - \ln \mathcal{P}_{\hat{x}'_{(j;Z),E}}\left(e^{i\theta_{\varphi_{U_k,j}}}\right) - 1\right) d\theta_{\varphi_{U_k,j}}\right)\right),
\end{aligned}$$

*where $P\left(\mathcal{M}_{U_k}\right)$ is the statistical private classical information of $U_k$, $\mathcal{X}_{AB}$ is the Holevo quantity of Bob's output, $\mathcal{X}_{BE}$ is the information leaked to the Eve in a reverse reconciliation, $\rho_k^{AB}$ is*

Bob's optimal output state, $\rho_k^{BE}$ is Eve's optimal state, $\sigma^{AB}$ is Bob's optimal output average state, $\sigma^{BE}$ is Eve's average state.

*Proof.*

The proof assumes the utilization of $m$ subcarriers for $\mathcal{M}_{U_k}$, and is demonstrated for an optimal ($j$-th) Gaussian quadrature component $\hat{x}_{j,U_k}$ of $U_k$. The results of the proof follow for the direct GQI method.

Without loss of generality, we derive the $S(\mathcal{M}_{U_k})$ statistical secret key rate via the $P(\mathcal{M}_{U_k})$ private classical information of $\mathcal{M}_{U_k}$, $S(\mathcal{M}_{U_k}) \leq \lim_{n \to \infty} \frac{1}{n} P(\mathcal{M}_{U_k})$. The logical channel between Alice ($A$) and Bob ($B$) is referred as $\mathcal{N}_{AB}$, while the logical channel between Bob and Eve ($E$) is denoted by $\mathcal{N}_{BE}$, therefore

$$P(\mathcal{M}_{U_k}) = \max_{\forall\, p_i, \rho_i} (\chi_{AB} - \chi_{BE}), \tag{138}$$

where

$$\chi_{AB} = S(\mathcal{N}_{AB}(\rho_{AB})) - \sum_i p_i S(\mathcal{N}_{AB}(\rho_i)) \tag{139}$$

and

$$\chi_{BE} = S(\mathcal{N}_{BE}(\rho_{BE})) - \sum_i p_i S(\mathcal{N}_{BE}(\rho_i)) \tag{140}$$

are the Holevo quantities between Alice and Bob, and Bob and Eve, $S(\rho) = -Tr(\rho \log(\rho))$ is the von Neumann entropy, while $\rho_{AB} = \sum_i p_i \rho_i$ and $\rho_{BE} = \sum_i p_i \rho_i$.

(Note: (138), in fact, is an oversimplified formula and cannot be considered as a general case. On the other hand, since the proof can be extended to an arbitrary multicarrier CVQKD setting, we further use this formula in the remaining parts. For the rigorous proofs on the various correlation measure formulas of a multicarrier CVQKD, see [5]. The direct reconciliation case also follows from the results by considering the logical channel $\mathcal{N}_{AE}$ of Alice and Eve, instead of $\mathcal{N}_{BE}$).

Thus, $P(\mathcal{M}_{U_k})$ at a reverse reconciliation is evaluated as

$$\begin{aligned} P(\mathcal{M}_{U_k}) = \max_{\forall\, p_i, \rho_i} & S\left(\mathcal{N}_{AB}\left(\sum_i p_i(\rho_i)\right)\right) - \sum_i p_i S(\mathcal{N}_{AB}(\rho_i)) \\ & - S\left(\mathcal{N}_{BE}\left(\sum_i p_i(\rho_i)\right)\right) + \sum_i p_i S(\mathcal{N}_{BE}(\rho_i)), \end{aligned} \tag{141}$$

where $\mathcal{N}(\rho_i)$ represents the $i$-th output density matrix.

Specifically, the $D(\cdot\|\cdot)$ quantum relative entropy function between density matrices $\rho$ and $\sigma$ is as

$$\begin{aligned} D(\rho\|\sigma) &= Tr(\rho \log(\rho)) - Tr(\rho \log(\sigma)) \\ &= Tr[\rho(\log(\rho) - \log(\sigma))]. \end{aligned} \tag{142}$$

The Holevo quantity can be expressed by the quantum relative entropy function as [39-40]

$$\chi = D(\rho_k \| \sigma), \tag{143}$$

where $\rho_k$ denotes an optimal (for which the Holevo quantity will be maximal) channel output state and $\sigma = \sum p_k \rho_k$.

The Holevo information $\mathcal{X}$ can be derived in terms of $D(\cdot \| \cdot)$ as

$$\begin{aligned}
\sum_k p_k D(\rho_k \| \sigma) &= \sum_k \left( p_k Tr(\rho_k \log(\rho_k)) - p_k Tr(\rho_k \log(\sigma)) \right) \\
&= \sum_k \left( p_k Tr(\rho_k \log(\rho_k)) \right) - Tr\left[ \sum_k (p_k \rho_k \log(\sigma)) \right] \\
&= \sum_k \left( p_k Tr(\rho_k \log(\rho_k)) \right) - Tr(\sigma \log(\sigma)) \\
&= S(\sigma) - \sum_k p_k S(\rho_k) = \mathcal{X}.
\end{aligned} \tag{144}$$

Therefore, $\mathcal{X}_{AB}$ can be rewritten as

$$\mathcal{X}_{AB} = S\left( \mathcal{N}_{AB}\left[ \sum_i p_i \rho_i \right] \right) - \sum_i p_i S(\mathcal{N}_{AB}(\rho_i)) = D(\rho_k^{AB} \| \sigma^{AB}). \tag{145}$$

The quantity $\mathcal{X}_{BE}$ measures the Holevo information which is leaked to Eve from Bob during a reverse reconciliation as

$$\mathcal{X}_{BE} = S\left( \mathcal{N}_{BE}\left[ \sum_i p_i \rho_i \right] \right) - \sum_i p_i S(\mathcal{N}_{BE}(\rho_i)) = D(\rho_k^{BE} \| \sigma^{BE}). \tag{146}$$

Using (145) and (146), $P(\mathcal{M}_{U_k})$ can be expressed via $D(\cdot \| \cdot)$ as

$$\begin{aligned}
P(\mathcal{M}_{U_k}) &= \min_\sigma \max_\rho D(\rho_k^{AB} \| \sigma^{AB}) - \min_\sigma \max_\rho D(\rho_k^{BE} \| \sigma^{BE}) \\
&= \min_\sigma \max_\rho D(\rho_k^{AB-BE} \| \sigma^{AB-BE}),
\end{aligned} \tag{147}$$

where $\rho_k^{AB-BE}$ is the final optimal density matrix, while $\sigma^{AB-BE}$ refers to the final output average density matrix.

Without loss of generality, let refer $\rho_k^{AB}$ to Bob's optimal $\hat{x}'_{j,U_k}$, and let $\rho_k^{BE}$ refer to Eve's optimal variable, $\hat{x}'_{j,E}$. Since $\hat{x}_{j,U_k}$, $\hat{x}'_{j,U_k}$, and $\hat{x}'_{j,E}$ are Gaussian random variables, it allows to express $D(\rho_k^{AB} \| \sigma^{AB})$ and $D(\rho_k^{BE} \| \sigma^{BE})$ from (145) and (146) via the (classical) relative entropy function $D_{AB}(\hat{x}'_{j,U_k} \| \hat{x}_{j,U_k})$ and $D_{BE}(\hat{x}'_{j,E} \| \hat{x}'_{j,U_k})$, as

$$D_{AB}(\hat{x}'_{j,U_k} \| \hat{x}_{j,U_k}) = \frac{1}{4\pi}\left( \int_{-\pi}^{\pi} \left( \frac{\mathcal{P}_{\hat{x}'_{j,U_k}}\left(e^{i\theta_{\varphi_{U_k,j}}}\right)}{\mathcal{P}_{\hat{x}_{j,U_k}}\left(e^{i\theta_{\varphi_{U_k,j}}}\right)} - \ln\frac{\mathcal{P}_{\hat{x}'_{j,U_k}}\left(e^{i\theta_{\varphi_{U_k,j}}}\right)}{\mathcal{P}_{\hat{x}_{j,U_k}}\left(e^{i\theta_{\varphi_{U_k,j}}}\right)} - 1 \right) d\theta_{\varphi_{U_k,j}} \right), \tag{148}$$

and

$$D_{BE}(\hat{x}'_{j,E} \| \hat{x}'_{j,U_k}) = \frac{1}{4\pi}\left( \int_{-\pi}^{\pi} \left( \frac{\mathcal{P}_{\hat{x}'_{j,E}}\left(e^{i\theta_{\varphi_{U_k,j}}}\right)}{\mathcal{P}_{\hat{x}'_{j,U_k}}\left(e^{i\theta_{\varphi_{U_k,j}}}\right)} - \ln\frac{\mathcal{P}_{\hat{x}'_{j,E}}\left(e^{i\theta_{\varphi_{U_k,j}}}\right)}{\mathcal{P}_{\hat{x}'_{j,U_k}}\left(e^{i\theta_{\varphi_{U_k,j}}}\right)} - 1 \right) d\theta_{\varphi_{U_k,j}} \right). \tag{149}$$

Hence, the formulas of (148) and (149) quantify the statistical information contained in $\hat{x}'_{j,U_k}$ about $\hat{x}_{j,U_k}$, and in $\hat{x}'_{j,E}$ about $\hat{x}'_{j,U_k}$, respectively.

Therefore,

$$D_{AB}\left(\hat{x}'_{j,U_k} \| \hat{x}_{j,U_k}\right) - D_{BE}\left(\hat{x}'_{j,E} \| \hat{x}'_{j,U_k}\right)$$
$$= \frac{1}{4\pi}\left(\int_{-\pi}^{\pi}\left(\frac{\mathcal{P}_{\hat{x}'_{j,U_k}}\left(e^{i\theta_{\varphi_{U_k,j}}}\right)}{\mathcal{P}_{\hat{x}_{j,U_k}}\left(e^{i\theta_{\varphi_{U_k,j}}}\right)} - \ln \frac{\mathcal{P}_{\hat{x}'_{j,U_k}}\left(e^{i\theta_{\varphi_{U_k,j}}}\right)}{\mathcal{P}_{\hat{x}_{j,U_k}}\left(e^{i\theta_{\varphi_{U_k,j}}}\right)} - 1\right)d\theta_{\varphi_{U_k,j}} - \int_{-\pi}^{\pi}\left(\frac{\mathcal{P}_{\hat{x}'_{j,E}}\left(e^{i\theta_{\varphi_{U_k,j}}}\right)}{\mathcal{P}_{\hat{x}'_{j,U_k}}\left(e^{i\theta_{\varphi_{U_k,j}}}\right)} - \ln \frac{\mathcal{P}_{\hat{x}'_{j,E}}\left(e^{i\theta_{\varphi_{U_k,j}}}\right)}{\mathcal{P}_{\hat{x}'_{j,U_k}}\left(e^{i\theta_{\varphi_{U_k,j}}}\right)} - 1\right)d\theta_{\varphi_{U_k,j}}\right). \tag{150}$$

Specifically, since Alice's $\hat{x}_{j,U_k}$ is unknown for Bob, and similarly, Bob's $\hat{x}'_{j,U_k}$ is unknown for Eve, allows us to rewrite $\mathcal{P}_{\hat{x}_{j,U_k}}\left(e^{i\theta_{\varphi_{U_k,j}}}\right)$ and $\mathcal{P}_{\hat{x}'_{j,U_k}}\left(e^{i\theta_{\varphi_{U_k,j}}}\right)$ of $\hat{x}_{j,U_k}$ and $\hat{x}'_{j,U_k}$ as

$$\mathcal{P}_{\hat{x}_{j,U_k}}\left(e^{i\theta_{\varphi_{U_k,j}}}\right) = \mathcal{P}_{\hat{x}'_{j,U_k}}\left(e^{i\theta_{\varphi_{U_k,j}}}\right) = 1. \tag{151}$$

The result in (151) is rooted in the fundamentals of the maximum entropy principle, e.g., $\mathcal{P}_{\hat{x}_{j,U_k}}\left(e^{i\theta_{\varphi_{U_k,j}}}\right)$, $\mathcal{P}_{\hat{x}'_{j,U_k}}\left(e^{i\theta_{\varphi_{U_k,j}}}\right)$ of $\hat{x}_{j,U_k}$ and $\hat{x}'_{j,U_k}$ are, in fact, white (constant) spectras with unit variance [28].

Let refer sub-index $(j;Z)$ to the optimal single-carrier quadrature variable, at $Z$ autocorrelation coefficients, obtained at $m$ subcarriers.

Then, using (150) and (151), $S\left(\mathcal{M}_{U_k}\right)$ at $n \to \infty$, can be expressed as a statistical information

$$\begin{aligned}S\left(\mathcal{M}_{U_k}\right) &\leq \lim_{n\to\infty}\frac{1}{n}P\left(\mathcal{M}_{U_k}\right) \\ &= \lim_{n\to\infty}\frac{1}{n}\lim_{Z\to\infty}\max_{\forall x}\left(D_{AB}\left(\hat{x}'_{(j;Z),U_k}\|\hat{x}_{(j;Z),U_k}\right) - D_{BE}\left(\hat{x}'_{(j;Z),E}\|\hat{x}'_{(j;Z),U_k}\right)\right) \\ &= \lim_{n\to\infty}\frac{1}{n}\lim_{Z\to\infty}\max_{\forall x}\left[\frac{1}{4\pi}\int_{-\pi}^{\pi}\left(\mathcal{P}_{\hat{x}'_{(j;Z),U_k}}\left(e^{i\theta_{\varphi_{U_k,j}}}\right) - \ln \mathcal{P}_{\hat{x}'_{(j;Z),U_k}}\left(e^{i\theta_{\varphi_{U_k,j}}}\right) - 1\right)d\theta_{\varphi_{U_k,j}} \right. \\ &\quad \left. -\left(\int_{-\pi}^{\pi}\left(\mathcal{P}_{\hat{x}'_{(j;Z),E}}\left(e^{i\theta_{\varphi_{U_k,j}}}\right) - \ln \mathcal{P}_{\hat{x}'_{(j;Z),E}}\left(e^{i\theta_{\varphi_{U_k,j}}}\right) - 1\right)d\theta_{\varphi_{U_k,j}}\right)\right],\end{aligned} \tag{152}$$

where

$$\begin{aligned}\lim_{Z\to\infty}\left(D_{AB}\left(\hat{x}'_{(j;Z),U_k}\|\hat{x}_{(j;Z),U_k}\right)\right) &= H\left(\hat{x}_{j,U_k}\right) - \lim_{Z\to\infty}H\left(\hat{x}'_{(j;Z),U_k}\right) \\ &= \tfrac{1}{2}\ln 2\pi + \tfrac{1}{2} + \tfrac{1}{4\pi}\int_{-\pi}^{\pi}\ln \mathcal{P}_{\hat{x}_{j,U_k}}\left(e^{i\theta_{\varphi_{U_k,j}}}\right)d\theta_{\varphi_{U_k,j}} \\ &\quad -\left(\lim_{Z\to\infty}\tfrac{1}{2}\ln 2\pi + \tfrac{1}{2} + \tfrac{1}{4\pi}\int_{-\pi}^{\pi}\ln \mathcal{P}_{\hat{x}'_{(j;Z),U_k}}\left(e^{i\theta_{\varphi_{U_k,j}}}\right)d\theta_{\varphi_{U_k,j}}\right),\end{aligned} \tag{153}$$

and

$$\lim_{Z \to \infty} \left( D_{BE} \left( \hat{x}'_{(j;Z),E} \| \hat{x}'_{(j;Z),U_k} \right) \right) = H \left( \hat{x}'_{j,U_k} \right) - \lim_{Z \to \infty} H \left( \hat{x}'_{(j;Z),E} \right)$$

$$= \tfrac{1}{2} \ln 2\pi + \tfrac{1}{2} + \tfrac{1}{4\pi} \int_{-\pi}^{\pi} \ln \mathcal{P}_{\hat{x}'_{j,U_k}} \left( e^{i\theta_{\varphi_{U_k,j}}} \right) d\theta_{\varphi_{U_k,j}} \quad (154)$$

$$- \left( \lim_{Z \to \infty} \tfrac{1}{2} \ln 2\pi + \tfrac{1}{2} + \tfrac{1}{4\pi} \int_{-\pi}^{\pi} \ln \mathcal{P}_{\hat{x}'_{(j;Z),E}} \left( e^{i\theta_{\varphi_{U_k,j}}} \right) d\theta_{\varphi_{U_k,j}} \right),$$

allowing to rewrite (152) as

$$S\left(\mathcal{M}_{U_k}\right) \leq \lim_{n \to \infty} \frac{1}{n} \max_{\forall x} \left( H\left(\hat{x}_{j,U_k}\right) - \lim_{Z \to \infty} H\left(\hat{x}'_{(j;Z),U_k}\right) - \left( H\left(\hat{x}'_{j,U_k}\right) - \lim_{Z \to \infty} H\left(\hat{x}'_{(j;Z),E}\right) \right) \right)$$

$$= \lim_{n \to \infty} \frac{1}{n} \max_{\forall x} \left( \left( H\left(\hat{x}_{j,U_k}\right) - H\left(\hat{x}'_{j,U_k}\right) \right) - \left( \lim_{Z \to \infty} H\left(\hat{x}'_{(j;Z),U_k}\right) - \lim_{Z \to \infty} H\left(\hat{x}'_{(j;Z),E}\right) \right) \right). \quad (155)$$

In particular, $S\left(\mathcal{M}_{U_k}\right)$ characterizes the statistical secret key rate as an entropy reduction; however, since it is determined via statistical functions, the $\mathcal{P}$ spectral densities used in (152) are, in fact, inferred.

Precisely, to verify (152), it is enough to prove that the limit exists, i.e., the quantities $D_{AB}\left(\hat{x}'_{(j;Z),U_k} \| \hat{x}_{(j;Z),U_k}\right)$ and $D_{BE}\left(\hat{x}'_{(j;Z),E} \| \hat{x}'_{(j;Z),U_k}\right)$ are bounded from above and non-decreasing in $Z$.

Without loss of generality, let $\mathcal{K}$ the set of all $\mathcal{P}$ spectras, associated with white (constant) spectras with unit variance. Let $Z$ the number of autocorrelation coefficients $\mathcal{A}_{\hat{x}'_{U_k,i}}(\cdot)$ and $\mathcal{A}_{\hat{x}'_{U_k,E}}(\cdot)$ obtained from the $m$ subcarriers at Bob and Eve, respectively, as

$$\mathcal{A}_{\hat{x}'_{U_k,i}}(l), \{l \in 0,...,Z-1\} \quad (156)$$

$$\mathcal{A}_{\hat{x}'_{U_k,E}}(l), (l), \{l \in 0,...,Z-1\}. \quad (157)$$

Then, let

$$\mathcal{B}_1(j;Z) \subset \mathcal{K} \quad (158)$$

and

$$\mathcal{B}_2(j;Z) \subset \mathcal{K} \quad (159)$$

be the sets of those spectra in $\mathcal{K}$ which are consistent with (156) and (157).

Specifically, let $\mathcal{P}_{\hat{x}'_{(j;Z),U_k}}\left(e^{i\theta_{\varphi_{U_k,j}}}\right) \in \mathcal{B}_1(j;Z)$ and $\mathcal{P}_{\hat{x}'_{(j;Z),E}}\left(e^{i\theta_{\varphi_{U_k,j}}}\right) \in \mathcal{B}_2(j;Z)$ be the unique power spectras of Bob and Eve, satisfying the conditions of Theorem 1.

Then, by some fundamental theory,

$$\mathcal{C}_{1,Z+1} \subset \mathcal{C}_{1,Z} \quad (160)$$

$$\mathcal{C}_{2,Z+1} \subset \mathcal{C}_{2,Z}, \quad (161)$$

where $\mathcal{C}_{1,Z}, \mathcal{C}_{1,Z+1}, \mathcal{C}_{2,Z}, \mathcal{C}_{2,Z+1}$ are constraint sets, defined as

$$\mathcal{C}_{1,Z+1} : \left(\mathcal{B}_1(j;Z+1) \cap \mathcal{K}\right), \mathcal{C}_{1,Z} : \left(\mathcal{B}_1(j;Z) \cap \mathcal{K}\right) \quad (162)$$

$$\mathcal{C}_{2,Z+1} : \left(\mathcal{B}_2(j;Z+1) \cap \mathcal{K}\right), \mathcal{C}_{2,Z} : \left(\mathcal{B}_2(j;Z) \cap \mathcal{K}\right), \quad (163)$$

such that for $\forall Z$

$$\mathcal{P}_{\hat{x}'_{j,U_k}}\left(e^{\mathrm{i}\theta_{\varphi_{U_k,j}}}\right) \subset \mathcal{C}_{1,Z}, \tag{164}$$

$$\mathcal{P}_{\hat{x}'_{j,E}}\left(e^{\mathrm{i}\theta_{\varphi_{U_k,j}}}\right) \subset \mathcal{C}_{2,Z}. \tag{165}$$

Thus, $\mathcal{P}_{\hat{x}'_{(j;Z),U_k}}\left(e^{\mathrm{i}\theta_{\varphi_{U_k,j}}}\right), \mathcal{P}_{\hat{x}'_{(j;Z+1),U_k}}\left(e^{\mathrm{i}\theta_{\varphi_{U_k,j}}}\right)$ and $\mathcal{P}_{\hat{x}'_{(j;Z),E}}\left(e^{\mathrm{i}\theta_{\varphi_{U_k,j}}}\right), \mathcal{P}_{\hat{x}'_{(j;Z+1),E}}\left(e^{\mathrm{i}\theta_{\varphi_{U_k,j}}}\right)$ have a maximal entropy with respect to the constraint sets $\mathcal{C}_{1,Z}, \mathcal{C}_{1,Z+1}, \mathcal{C}_{2,Z}, \mathcal{C}_{2,Z+1}$.

Specifically,
$$H\left(\mathcal{P}_{\hat{x}'_{(j;Z),U_k}}\left(e^{\mathrm{i}\theta_{\varphi_{U_k,j}}}\right)\right) \geq H\left(\mathcal{P}_{\hat{x}'_{(j;Z+1),U_k}}\left(e^{\mathrm{i}\theta_{\varphi_{U_k,j}}}\right)\right) \geq H\left(\mathcal{P}_{\hat{x}'_{j,U_k}}\left(e^{\mathrm{i}\theta_{\varphi_{U_k,j}}}\right)\right), \tag{166}$$

and
$$H\left(\mathcal{P}_{\hat{x}'_{(j;Z),E}}\left(e^{\mathrm{i}\theta_{\varphi_{U_k,j}}}\right)\right) \geq H\left(\mathcal{P}_{\hat{x}'_{(j;Z+1),E}}\left(e^{\mathrm{i}\theta_{\varphi_{U_k,j}}}\right)\right) \geq H\left(\mathcal{P}_{\hat{x}'_{j,E}}\left(e^{\mathrm{i}\theta_{\varphi_{U_k,j}}}\right)\right), \tag{167}$$

where
$$H(x) = \tfrac{1}{2}\ln 2\pi + \tfrac{1}{2} + \tfrac{1}{4\pi}\int_{-\pi}^{\pi}\ln \mathcal{P}_x\left(e^{\mathrm{i}\theta_{\varphi_{U_k,j}}}\right)d\theta_{\varphi_{U_k,j}}. \tag{168}$$

In particular, $D_X(x\|\tilde{x})$ for an $x$ Gaussian random variable can be expressed as
$$D_X(x\|\tilde{x}) = c - H(x), \tag{169}$$

where $\tilde{x}$ is a reference variable, and $c$ is a constant, evaluated as
$$c = \tfrac{1}{2}\ln 2\pi + \tfrac{1}{2}, \tag{170}$$

Thus, the following relation brings up between Alice and Bob,
$$\begin{aligned}D_{AB}\left(\mathcal{P}_{\hat{x}'_{(j;Z),U_k}}\left(e^{\mathrm{i}\theta_{\varphi_{U_k,j}}}\right)\Big\|\mathcal{P}_{\hat{x}_{j,U_k}}\left(e^{\mathrm{i}\theta_{\varphi_{U_k,j}}}\right)\right) &\leq D_{AB}\left(\mathcal{P}_{\hat{x}'_{(j;Z+1),U_k}}\left(e^{\mathrm{i}\theta_{\varphi_{U_k,j}}}\right)\Big\|\mathcal{P}_{\hat{x}_{j,U_k}}\left(e^{\mathrm{i}\theta_{\varphi_{U_k,j}}}\right)\right)\\ &\leq D_{AB}\left(\mathcal{P}_{\hat{x}'_{j,U_k}}\left(e^{\mathrm{i}\theta_{\varphi_{U_k,j}}}\right)\Big\|\mathcal{P}_{\hat{x}_{j,U_k}}\left(e^{\mathrm{i}\theta_{\varphi_{U_k,j}}}\right)\right),\end{aligned} \tag{171}$$

and between Bob and Eve,
$$\begin{aligned}D_{BE}\left(\mathcal{P}_{\hat{x}'_{(j;Z),E}}\left(e^{\mathrm{i}\theta_{\varphi_{U_k,j}}}\right)\Big\|\mathcal{P}_{\hat{x}'_{j,U_k}}\left(e^{\mathrm{i}\theta_{\varphi_{U_k,j}}}\right)\right) &\leq D_{BE}\left(\mathcal{P}_{\hat{x}'_{(j;Z+1),E}}\left(e^{\mathrm{i}\theta_{\varphi_{U_k,j}}}\right)\Big\|\mathcal{P}_{\hat{x}'_{j,U_k}}\left(e^{\mathrm{i}\theta_{\varphi_{U_k,j}}}\right)\right)\\ &\leq D_{BE}\left(\mathcal{P}_{\hat{x}'_{j,E}}\left(e^{\mathrm{i}\theta_{\varphi_{U_k,j}}}\right)\Big\|\mathcal{P}_{\hat{x}'_{j,U_k}}\left(e^{\mathrm{i}\theta_{\varphi_{U_k,j}}}\right)\right),\end{aligned} \tag{172}$$

thus the limit in (152) is immediately concluded.

∎

**Lemma 1**. *Increasing the number $m$ of subcarriers of $\mathcal{M}_{U_k}$, increases $S(\mathcal{M}_{U_k})$.*

*Proof.*
Let sub-index $(j, Z, m)$ refer to the $j$-th optimal single-carrier at $Z$ autocorrelation coefficients and $m$ subcarriers.
At $m+1$ subcarriers of $\mathcal{M}_{U_k}$,

$$D_{AB}\left(\mathcal{P}_{\hat{x}'_{(j;Z,m),U_k}}\left(e^{\mathrm{i}\theta_{\varphi_{U_k,j}}}\right)\middle\|\mathcal{P}_{\hat{x}_{j,U_k}}\left(e^{\mathrm{i}\theta_{\varphi_{U_k,j}}}\right)\right) \leq D_{AB}\left(\mathcal{P}_{\hat{x}'_{(j;Z,m+1),U_k}}\left(e^{\mathrm{i}\theta_{\varphi_{U_k,j}}}\right)\middle\|\mathcal{P}_{\hat{x}_{j,U_k}}\left(e^{\mathrm{i}\theta_{\varphi_{U_k,j}}}\right)\right)$$
$$\leq D_{AB}\left(\mathcal{P}_{\hat{x}'_{j,U_k}}\left(e^{\mathrm{i}\theta_{\varphi_{U_k,j}}}\right)\middle\|\mathcal{P}_{\hat{x}_{j,U_k}}\left(e^{\mathrm{i}\theta_{\varphi_{U_k,j}}}\right)\right), \quad (173)$$

and

$$H\left(\mathcal{P}_{\hat{x}'_{(j;Z,m),U_k}}\left(e^{\mathrm{i}\theta_{\varphi_{U_k,j}}}\right)\right) \geq H\left(\mathcal{P}_{\hat{x}'_{(j;Z,m+1),U_k}}\left(e^{\mathrm{i}\theta_{\varphi_{U_k,j}}}\right)\right) \geq H\left(\mathcal{P}_{\hat{x}'_{j,U_k}}\left(e^{\mathrm{i}\theta_{\varphi_{U_k,j}}}\right)\right). \quad (174)$$

Precisely, from the security thresholds of multicarrier CVQKD [2], [5], follows that

$$\chi_{AB}^{(m+1)} - \chi_{AB}^{(m)} \geq \chi_{BE}^{(m+1)} - \chi_{BE}^{(m)}, \quad (175)$$

where $\chi_{AB}^{(m)}$, $\chi_{BE}^{(m)}$ and $\chi_{AB}^{(m+1)}$, $\chi_{BE}^{(m+1)}$ are the corresponding Holevo quantities at $m$ and $m+1$ subcarriers of $\mathcal{M}_{U_k}$, therefore at $Z \to \infty$

$$S\left(\mathcal{M}_{U_k}^{(m)}\right) \leq S\left(\mathcal{M}_{U_k}^{(m+1)}\right) \leq \lim_{n\to\infty} \frac{1}{n} \max_{\forall x} D_{AB}\left(\mathcal{P}_{\hat{x}'_{j,U_k}}\left(e^{\mathrm{i}\theta_{\varphi_{U_k,j}}}\right)\middle\|\mathcal{P}_{\hat{x}_{j,U_k}}\left(e^{\mathrm{i}\theta_{\varphi_{U_k,j}}}\right)\right), \quad (176)$$

where $\mathcal{M}_{U_k}^{(m)}$ and $\mathcal{M}_{U_k}^{(m+1)}$ refer to the logical channel of $U_k$ at $m$ and $m+1$ subcarriers, respectively.

Without loss of generality, let $\mathcal{V}$ be a poset (partially ordered set) of the $\mathcal{N}_{U_k,i}, i = 0,...,m-1$, $m$ sub-channel outputs of $\mathcal{M}_{U_k}^{(m)}$, and let $\mathrm{Y} \in \mathcal{V}$ be an output realization.

Let

$$\mathbf{S}_{\mathrm{Y}}\left(\mathcal{M}_{U_k}^{(m)}\right) = \left[S\left(\mathcal{M}_{U_k}^{(0)}\right),...,S\left(\mathcal{M}_{U_k}^{(m-1)}\right)\right]^T, \quad (177)$$

where $S\left(\mathcal{M}_{U_k}^{(i)}\right)$ is achieved at $i$ noisy subcarrier quadrature components $\hat{x}'_{h,U_k}, h = 0,...,i-1$.

Specifically, (177) provides a cumulative statistical secret key rate of the $m$ sub-channels, $\mathcal{N}_{U_k,i}$, of $\mathcal{M}_{U_k}$.

From a fundamental information scalability principle [28], it follows that if

$$\mathrm{Y}_1 \overset{\inf}{\geq} \mathrm{Y}_2, \ \mathrm{Y}_1 \in \mathcal{V}, \ \mathrm{Y}_2 \in \mathcal{V}, \quad (178)$$

where $\inf$ stands for the information scalability, then

$$\mathbf{S}_{\mathrm{Y}_1}\left(\mathcal{M}_{U_k}^{(m)}\right) \geq \mathbf{S}_{\mathrm{Y}_2}\left(\mathcal{M}_{U_k}^{(m)}\right). \quad (179)$$

Note, for an optimal $\tilde{\mathrm{Y}} \in \mathcal{V}$, $\tilde{\mathrm{Y}} \overset{\inf}{\geq} \mathrm{Y}, \forall \mathrm{Y} \in \mathcal{V}$.

∎

# 5 Conclusions

We defined a statistical quadrature inference method for multicarrier CVQKD. We introduced the frameworks of GQI and DGQI. The GQI provides a statistical estimation of the continuous variable Gaussian subcarrier quadratures from the observed noisy Gaussian subcarriers, conveyed via the Gaussian sub-channels. The DGQI direct method has a flexible realization with a low-complexity mathematical apparatus. We proved that a GQI-based quadrature inference allows to achieve a vanishing magnitude error in the quadrature estimation procedure. Using the statistical functions of GQI, we proved the statistical secret key rate formulas. The GQI and DGQI frameworks can be established in an arbitrary CVQKD setting, and are implementable by standard low-complexity functions, which is specifically convenient for an experimental CVQKD scenario.

# Acknowledgements

The author would like to thank Professor Sandor Imre for useful discussions. This work was partially supported by the GOP-1.1.1-11-2012-0092 (*Secure quantum key distribution between two units on optical fiber network*) project sponsored by the EU and European Structural Fund, by the Hungarian Scientific Research Fund - OTKA K-112125, and by the COST Action MP1006.

# Supplemental Information

## S.1 Notations

The notations of the manuscript are summarized in Table S.1.

**Table S.1.** Summary of notations.

| | |
|---|---|
| $i$ | Index for the $i$-th subcarrier Gaussian CV, $\left|\phi_i\right\rangle = x_i + \mathrm{i}p_i$. |
| $j$ | Index for the $j$-th Gaussian single-carrier CV, $\left|\varphi_j\right\rangle = x_j + \mathrm{i}p_j$. |
| $l$ | Number of Gaussian sub-channels $\mathcal{N}_i$ for the transmission of the Gaussian subcarriers. The overall number of the sub-channels is $n$. The remaining $n-l$ sub-channels do not transmit valuable information. |
| $x_i, p_i$ | Position and momentum quadratures of the $i$-th Gaussian subcarrier, $\left|\phi_i\right\rangle = x_i + \mathrm{i}p_i$. |
| $x'_i, p'_i$ | Noisy position and momentum quadratures of Bob's $i$-th noisy subcarrier Gaussian CV, $\left|\phi'_i\right\rangle = x'_i + \mathrm{i}p'_i$. |
| $x_j, p_j$ | Position and momentum quadratures of the $j$-th Gaussian single-carrier $\left|\varphi_j\right\rangle = x_j + \mathrm{i}p_j$. |
| $x'_j, p'_j$ | Noisy position and momentum quadratures of Bob's $j$-th recovered single-carrier Gaussian CV $\left|\varphi'_j\right\rangle = x'_j + \mathrm{i}p'_j$. |
| $x_{A,i}, p_{A,i}$ | Alice's quadratures in the transmission of the $i$-th subcarrier. |
| $\left|\phi_i\right\rangle, \left|\phi'_i\right\rangle$ | Transmitted and received Gaussian subcarriers. |
| $\mathbf{z} \in \mathcal{CN}\left(0, \mathbf{K_z}\right)$ | A $d$-dimensional input CV vector to transmit valuable information. |
| $\mathbf{z}'^T$ | A $d$-dimensional noisy output vector, $\mathbf{z}'^T = \mathbf{A}^\dagger \mathbf{z} + \left(F^d(\Delta)\right)^T = (z'_0, ..., z'_{d-1})$, where $z'_j = \left(\frac{1}{l}\sum_{i=0}^{l-1} F(T_{j,i}(\mathcal{N}_{j,i}))\right)z_j + F(\Delta) \in \mathcal{CN}\left(0, 2\left(\sigma^2_{\omega_0} + \sigma^2_{\mathcal{N}}\right)\right)$. |
| $x(n)$ | A WSS (wide-sense stationary) process. |
| $\mathcal{A}_{x(n)}(\cdot)$ | Autocorrelation function (sequence) of $x(n)$. |
| $Z$ | Number of autocorrelation coefficients. |
| $\mathbf{C}_{xx}$ | An $n \times n$ covariance matrix associated with $x(n)$. |
| $\mathcal{P}_x\left(e^{i\omega}\right)$ | Power spectrum of $x(n)$, evaluated as $$\mathcal{P}_x\left(e^{i\omega}\right) = \sum_{l=-\infty}^{+\infty} \mathcal{A}_{x(n)}(l) e^{-i\omega l}, \text{ where } \omega \in [-\pi, \pi].$$ |
| $\mathcal{S}_x\left(e^{i\omega}\right)$ | Spectral density of $x(n)$. |
| $H(x)$ | Entropy rate of $x(n)$. |
| $f_E(x)$ | Empirical density function. |
| $D_X(\cdot)$ | Classical relative entropy function. |
| $D(\cdot\|\cdot)$ | Quantum relative entropy function, $$D(\rho\|\sigma) = Tr(\rho \log(\rho)) - Tr(\rho \log(\sigma)) \\ = Tr[\rho(\log(\rho) - \log(\sigma))],$$ where $\rho$ and $\sigma$ are density matrices. |
| $S(\rho)$ | Neumann entropy, $S(\rho) = -Tr(\rho \log(\rho))$. |
| $N_n(f_E(x))$ | Number of $n$-tuples $(x_1, ..., x_n) \in X^n$ with a given empirical density $f_E(x)$. |
| $\mathfrak{L}^p(a,b)$ | Space of Lebesgue-measurable functions. |
| $\mathcal{H}_f(a,b)$ | Functional Hilbert space, $a=-\pi, b=\pi$, $$\mathcal{H}_f(-\pi,\pi) = \mathcal{L}^2(-\pi,\pi),$$ with a norm $$\|f\|_2 = \tfrac{1}{2\pi}\sqrt{\int_{-\pi}^{\pi}|f(x)|^2\, dx}\, .$$ |
| $\mathcal{M}_{U_k}$ | Logical channel of user $U_k, k=0,\ldots,K-1$, where $K$ is the number of total users, $$\mathcal{M}_{U_k} = \left[\mathcal{N}_{U_k,0},\ldots,\mathcal{N}_{U_k,m-1}\right]^T,$$ and $\mathcal{N}_{U_k,i}$ is the $i$-th sub-channel of $\mathcal{M}_{U_k}$, $m$ is the number of subcarriers dedicated to $U_k$. |
| $E\!\left(U^{-1}\!\left(\varphi_{U_k,j}\right)\right)$ | Estimate of $U^{-1}\!\left(\varphi_{U_k,j}\right)$ where $E(\cdot)$ stands for the estimator function, $U^{-1}$ is the inverse CVQFT operation. |
| $E\!\left(U^{-1}\!\left(x_{U_k,j}\right)\right)$ | Estimate of $U^{-1}\!\left(x_{U_k,j}\right)$, where $x_{U_k,j}$ is the quadrature component of $\varphi_{U_k,j}$, where $E(\cdot)$ stands for the estimator function, $U^{-1}$ is the inverse CVQFT operation. |
| $M$ | Measurement operator, homodyne or heterodyne measurement. |
| $\theta_{\varphi_{U_k,j}}$ | $\theta_{\varphi_{U_k,j}} = \pi/\Omega$, where $\Omega = \sigma_{\omega_0}^2/\sigma_\omega^2$, and $\sigma_{\omega_0}^2$, $\sigma_\omega^2$ are the single-carrier and multicarrier modulation variances, $\sigma_\omega^2 \leq \sigma_{\omega_0}^2$, $\Omega \geq 1$, $\theta_{\varphi_{U_k,j}} \leq \pi$. |
| $\lambda$ | Lagrange multiplier, $$\lambda = \left|F(T_\mathcal{N}^*)\right|^2 = \tfrac{1}{l}\sum_{i=0}^{l-1}\left|F(T_i^*(\mathcal{N}_i))\right|^2$$ $$= \tfrac{1}{l}\sum_{i=0}^{l-1}\left|\sum_{k=0}^{l-1} T_k^* e^{\frac{-i2\pi ik}{n}}\right|^2,$$ where $T_\mathcal{N}^*$ is the expected transmittance of the $l$ sub-channels under an optimal Gaussian attack. |

| | |
|---|---|
| $\tilde{\lambda}_i$ | Optimal Lagrange multipliers. |
| $\mathcal{G}_i\left(e^{\mathrm{i}\theta_{\varphi_{U_k,j}}}\right)$ | $\mathcal{G}_i(x) = \mathcal{T}_i(x)\mathcal{T}_i(x^{-1})$, where $$\mathcal{T}_i\left(e^{\mathrm{i}\theta_{\varphi_{U_k,j}}}\right) = \begin{cases} T_i(\mathcal{N}_{U_k,i}), & \text{if } \left|\theta_{\varphi_{U_k,j}}\right| \leq \frac{\pi}{\Omega}, \\ 0, & \text{otherwise} \end{cases}$$ where $\mathcal{N}_{U_k,i}$ is the $i$-th Gaussian sub-channel of $\mathcal{M}_{U_k}$ of user $U_k$. |
| $\wp$ | Set, defined as $$\wp = \left\{ \tfrac{1}{2\pi} \int_{-\pi}^{\pi} \mathcal{P}\left(e^{\mathrm{i}\theta_{\varphi_{U_k,j}}}\right) \mathcal{G}_i\left(e^{\mathrm{i}\theta_{\varphi_{U_k,j}}}\right) e^{\mathrm{i}\Omega\theta_{\varphi_{U_k,j}}q} d\theta_{\varphi_{U_k,j}}, \right.$$ $$\forall x'_{U_k,i} \in \mathrm{X}_{\phi'_{U_k,i}}, q = 0,\ldots,L-1$$ $$\cup \mathcal{P}\left(e^{\mathrm{i}\theta_{\varphi_{U_k,j}}}\right) \in \mathfrak{L}^1(-\pi,\pi),$$ $$\left. \cup \mathcal{P}\left(e^{\mathrm{i}\theta_{\varphi_{U_k,j}}}\right) \geq 0 \right\},$$ |
| $\mathrm{X}_{\phi'_{U_k,i}}$ | Set of the autocorrelation functions, $$\mathrm{X}_{\phi'_{U_k,i}} = \left\{\mathcal{A}_{x'_{U_k,0}},\ldots,\mathcal{A}_{x'_{U_k,m-1}}\right\}.$$ |
| $\mathcal{F}_i(x)$ | Transfer function $$\mathcal{F}_i(x) = \sum_{q=-(L-1)}^{L-1} 2\lambda_{iq} x^{-q},$$ where $\lambda_{iq}$ are the Lagrange multipliers, $i = 0,\ldots,m-1, q = 0,\ldots,L-1$. |
| $\Theta$ | Constraint, $\Theta(\Gamma) = \sum_{i=0}^{m-1}\left(\omega_i(\Gamma) - \mathcal{A}_{x'_{U_k,i}}\right)^2$. |
| $\Gamma$ | Lagrangian set, $\Gamma = \left[\lambda_0,\ldots,\lambda_{(m-1)}\right]^T$. |
| $\mathcal{C}$ | Constraint set. |
| $\delta(\cdot)$ | Entropy measure function. |
| $H_B$ | Burg entropy, $\delta(x) = H_B(x) = -\log x$. |
| $\Upsilon$ | Plane surface. |
| $\mathfrak{T}$ | Closed triangular region in $\mathbb{R}^3$. |
| $*$ | Linear convolution operator. |
| $\mathfrak{f}$ | Function, provides an $\varepsilon$ minimal magnitude error, for $$E\left(U^{-1}\left(x_{U_k,j}\right)\right) = F^{-1}\left(x'_{U_k,j}\right) * F^{-1}\left(\beta_i\right), i = 0,...,m-1.$$ |
| $\varepsilon$ | Minimal magnitude error, $\varepsilon = \arg\min \varepsilon_{\max}$, where $\varepsilon_{\max} = \max_{\forall x}\left\| \left|x_{U_k,j}\right|^2 - \left|x'_{U_k,j}\right|^2 \right\|$, $x_{U_k,j}$, $x'_{U_k,j}$ are the input, output single-carrier quadratures, $F^{-1}(\cdot)$ is the inverse FFT operation. |
| $\mathbf{x}_{U_k,i}$ | Input subcarrier vector, $\mathbf{x}_{U_k,i} = \left(x_{U_k,0},...,x_{U_k,m-1}\right)^T$. |
| $\mathbf{x}'_{U_k,i}$ | Output subcarrier vector, $\mathbf{x}'_{U_k,i} = \left(x'_{U_k,0},...,x'_{U_k,m-1}\right)^T$. |
| $\mho_{\sigma_\omega^2}$ | Offset range of $\varepsilon_{\max}$, $\mho_{\sigma_\omega^2} : \left\{0 \leq \sigma_\omega^2 \leq 0.5\right\}$, quantifies the maximal deviation from 0 dB at $0 \leq \sigma_\omega^2 \leq 0.5$. |
| $\mathrm{A}_i$ | Operator, $\mathrm{A}_i = 2\sigma_\omega^2 \frac{i}{m}$, $i = 0,...,\frac{m}{2}$, where $\sigma_\omega^2$ is the subcarrier modulation variance, $m$ is the number of subcarriers of $U_k$. |
| $\mathfrak{E}$ | Estimator function, evaluated as $$\mathfrak{E} = \begin{cases} i = 0 : \mathfrak{E}\left(\mathrm{A}_0\right) = \frac{1}{m^2}\left|x'_{U_k,0}\right|^2 \\ i = 1,...,\frac{m}{2}-1 : \mathfrak{E}\left(\mathrm{A}_i\right) = \frac{1}{m^2}\left(\left|x'_{U_k,i}\right|^2 + \left|x'_{U_k,m-i}\right|^2\right), \\ i = \frac{m}{2} : \mathfrak{E}\left(\mathrm{A}_{\frac{m}{2}}\right) = \frac{1}{m^2}\left|x'_{U_k,\frac{m}{2}}\right|^2 \end{cases}$$ |

| | |
|---|---|
| $\alpha$, $\alpha_\varepsilon$ | $\alpha = m\sum_{i=0}^{m-1}\beta_i^2$, $\alpha_\varepsilon = m\sum_{z=0}^{m-1}\beta_{z,\varepsilon}^2$. |
| $\beta_i$ | Parameter, defined to evaluate function $\mathfrak{f}(s)$ as $$\mathfrak{f}(s) = \frac{1}{\alpha}\left|\sum_{i=0}^{m-1}\beta_i e^{\frac{i2\pi is}{m}}\right|^2 = \frac{1}{\alpha}\left|\int_{-m/2}^{m/2}\cos\left(\frac{2\pi si}{m}\right)\beta\left(i-\frac{m}{2}\right)di\right|^2.$$ |
| $\beta_{i,\varepsilon}$ | Function to minimize $\varepsilon$, $\beta_{i,\varepsilon} = 1 + \sum_{y=1}^{P}C_y\cos(yQ_i)$, where $C_0$ is arbitrarily set to unity, $P$ is the number of $C_y$ coefficients, while $Q_i = \frac{2\pi i}{m}, i = 0,\ldots,m-1$. |
| $\mathrm{N}(\cdot)$ | Normalization term. |
| $\mathrm{N}(\beta_i)$ | Normalization of $\beta_i$, $\mathrm{N}(\beta_i) = \beta_i\frac{m}{\sum_{i=0}^{m-1}\beta_i}$. |
| $\mathrm{N}(\beta_{i,\varepsilon})$ | Normalization of $\beta_{i,\varepsilon}$, $\mathrm{N}(\beta_{i,\varepsilon}) = \beta_{i,\varepsilon}\frac{m}{\sum_{i=0}^{m-1}\beta_{i,\varepsilon}}$. |
| $P(\mathcal{M}_{U_k})$ | Statistical private classical information of $U_k$. |
| $S(\mathcal{M}_{U_k})$ | Statistical secret key rate of $U_k$. |
| $\hat{x}_{(j:Z),U_k}$, $\hat{x}'_{(j:Z),U_k}$, $\hat{x}'_{(j:Z),E}$ | Optimal quadratures of Alice, Bob and Eve, obtained at $Z$ autocorrelation coefficients. |
| $\mathcal{X}_{AB}$ | Holevo quantity of Bob's output. |
| $\mathcal{X}_{BE}$ | Holevo information leaked to the Eve in a reverse reconciliation. |
| $\rho_k^{AB}$ | Bob's optimal output density matrix. |
| $\rho_k^{BE}$ | Eve's optimal density matrix, at a reverse reconciliation. |
| $\sigma^{AB}$ | Bob's optimal output average density matrix. |
| $\sigma^{BE}$ | Eve's average density matrix, at a reverse reconciliation. |
| $\mathcal{K}$ | Set of all $\mathcal{P}$ spectras, associated with white (constant) spectras with unit variance. |
| $\mathcal{B}_1, \mathcal{B}_2$ | Sets of spectras in $\mathcal{K}$, $\mathcal{B}_1(j;Z) \subset \mathcal{K}$, $\mathcal{B}_2(j;Z) \subset \mathcal{K}$. |
| $\mathcal{C}_{1,Z}, \mathcal{C}_{1,Z+1}, \mathcal{C}_{2,Z}, \mathcal{C}_{2,Z+1}$ | Constraint sets, defined as $$\mathcal{C}_{1,Z+1} : (\mathcal{B}_1(j;Z+1) \cap \mathcal{K}), \mathcal{C}_{1,Z} : (\mathcal{B}_1(j;Z) \cap \mathcal{K})$$ $$\mathcal{C}_{2,Z+1} : (\mathcal{B}_2(j;Z+1) \cap \mathcal{K}), \mathcal{C}_{2,Z} : (\mathcal{B}_2(j;Z) \cap \mathcal{K}).$$ |
| $\mathbf{S}_Y\left(\mathcal{M}_{U_k}^{(m)}\right)$ | Vector, provides a cumulative statistical secret key rate of the $m$ sub-channels, $\mathcal{N}_{U_k,i}$ of $\mathcal{M}_{U_k}$, $\mathbf{S}_Y\left(\mathcal{M}_{U_k}^{(m)}\right) = \left[S\left(\mathcal{M}_{U_k}^{(0)}\right),\ldots,S\left(\mathcal{M}_{U_k}^{(m-1)}\right)\right]^T$, where $S\left(\mathcal{M}_{U_k}^{(i)}\right)$ is the secret key rate obtained at $i$ subcarriers. |
| $\mathcal{V}$ | Poset (partially ordered set). |
| $Y \in \mathcal{V}$ | An output realization of $\mathcal{M}_{U_k}^{(m)}$. |
| inf | Information scalability. |
| $U_{K_{out}}$ | The unitary CVQFT operation, $U_{K_{out}} = \frac{1}{\sqrt{K_{out}}} e^{\frac{-i2\pi ik}{K_{out}}}$, $i,k = 0,\ldots,K_{out}-1$, $K_{out} \times K_{out}$ unitary matrix. |
| $U_{K_{in}}$ | The unitary inverse CVQFT operation, $U_{K_{in}} = \frac{1}{\sqrt{K_{in}}} e^{\frac{i2\pi ik}{K_{in}}}$, $i,k = 0,\ldots,K_{in}-1$, $K_{in} \times K_{in}$ unitary matrix. |
| $z \in \mathcal{CN}(0,\sigma_z^2)$ | The variable of a single-carrier Gaussian CV state, $\left|\varphi_i\right\rangle \in \mathcal{S}$. Zero-mean, circular symmetric complex Gaussian random variable, $\sigma_z^2 = \mathbb{E}\left[\left|z\right|^2\right] = 2\sigma_{\omega_0}^2$, with i.i.d. zero mean, Gaussian random quadrature components $x, p \in \mathbb{N}(0,\sigma_{\omega_0}^2)$, where $\sigma_{\omega_0}^2$ is the variance. |
| $\Delta \in \mathcal{CN}(0,\sigma_\Delta^2)$ | The noise variable of the Gaussian channel $\mathcal{N}$, with i.i.d. zero-mean, Gaussian random noise components on the posi- |
| | tion and momentum quadratures $\Delta_x, \Delta_p \in \mathbb{N}\left(0, \sigma_\mathcal{N}^2\right)$, $\sigma_\Delta^2 = \mathbb{E}\left[\left|\Delta\right|^2\right] = 2\sigma_\mathcal{N}^2$. |
| $d \in \mathcal{CN}\left(0, \sigma_d^2\right)$ | The variable of a Gaussian subcarrier CV state, $\left|\phi_i\right\rangle \in \mathcal{S}$. Zero-mean, circular symmetric Gaussian random variable, $\sigma_d^2 = \mathbb{E}\left[\left|d\right|^2\right] = 2\sigma_\omega^2$, with i.i.d. zero mean, Gaussian random quadrature components $x_d, p_d \in \mathbb{N}\left(0, \sigma_\omega^2\right)$, where $\sigma_\omega^2$ is the (constant) modulation variance of the Gaussian subcarrier CV state. |
| $F^{-1}\left(\cdot\right) = \mathrm{CVQFT}^\dagger\left(\cdot\right)$ | The inverse CVQFT transformation, applied by the encoder, continuous-variable unitary operation. |
| $F\left(\cdot\right) = \mathrm{CVQFT}\left(\cdot\right)$ | The CVQFT transformation, applied by the decoder, continuous-variable unitary operation. |
| $F^{-1}\left(\cdot\right) = \mathrm{IFFT}\left(\cdot\right)$ | Inverse FFT transform, applied by the encoder. |
| $\sigma_{\omega_0}^2$ | Single-carrier modulation variance. |
| $\sigma_\omega^2 = \tfrac{1}{l}\sum_l \sigma_{\omega_i}^2$ | Multicarrier modulation variance. Average modulation variance of the $l$ Gaussian sub-channels $\mathcal{N}_i$. |
| $\left|\phi_i\right\rangle = \left|\mathrm{IFFT}\left(z_{k,i}\right)\right\rangle$ $= \left|F^{-1}\left(z_{k,i}\right)\right\rangle = \left|d_i\right\rangle.$ | The $i$-th Gaussian subcarrier CV of user $U_k$, where IFFT stands for the Inverse Fast Fourier Transform, $\left|\phi_i\right\rangle \in \mathcal{S}$, $d_i \in \mathcal{CN}\left(0, \sigma_{d_i}^2\right)$, $\sigma_{d_i}^2 = \mathbb{E}\left[\left|d_i\right|^2\right]$, $d_i = x_{d_i} + \mathrm{i}p_{d_i}$, $x_{d_i} \in \mathbb{N}\left(0, \sigma_{\omega_F}^2\right)$, $p_{d_i} \in \mathbb{N}\left(0, \sigma_{\omega_F}^2\right)$ are i.i.d. zero-mean Gaussian random quadrature components, and $\sigma_{\omega_F}^2$ is the variance of the Fourier transformed Gaussian state. |
| $\left|\varphi_{k,i}\right\rangle = \mathrm{CVQFT}\left(\left|\phi_i\right\rangle\right)$ | The decoded single-carrier CV of user $U_k$ from the subcarrier CV, expressed as $F\left(\left|d_i\right\rangle\right) = \left|F\left(F^{-1}\left(z_{k,i}\right)\right)\right\rangle = \left|z_{k,i}\right\rangle$. |
| $\mathcal{N}$ | Gaussian quantum channel. |
| $\mathcal{N}_i, i = 0,\ldots,n-1$ | Gaussian sub-channels. |
| $T(\mathcal{N})$ | Channel transmittance, normalized complex random variable, $T(\mathcal{N}) = \operatorname{Re} T(\mathcal{N}) + \mathrm{i}\operatorname{Im} T(\mathcal{N}) \in \mathcal{C}$. The real part identifies the position quadrature transmission, the imaginary part identifies the transmittance of the position quadrature. |
| $T_i(\mathcal{N}_i)$ | Transmittance coefficient of Gaussian sub-channel $\mathcal{N}_i$, $T_i(\mathcal{N}_i) = \operatorname{Re}(T_i(\mathcal{N}_i)) + \mathrm{i}\operatorname{Im}(T_i(\mathcal{N}_i)) \in \mathcal{C}$, quantifies the position and momentum quadrature transmission, with (normalized) real and imaginary parts $0 \leq \operatorname{Re} T_i(\mathcal{N}_i) \leq 1/\sqrt{2}$, $0 \leq \operatorname{Im} T_i(\mathcal{N}_i) \leq 1/\sqrt{2}$, where $\operatorname{Re} T_i(\mathcal{N}_i) = \operatorname{Im} T_i(\mathcal{N}_i)$. |
| $T_{Eve}$ | Eve's transmittance, $T_{Eve} = 1 - T(\mathcal{N})$. |
| $T_{Eve,i}$ | Eve's transmittance for the $i$-th subcarrier CV. |
| $\mathbf{z} = \mathbf{x} + \mathrm{i}\mathbf{p} = (z_0,\ldots,z_{d-1})^T$ | A $d$-dimensional, zero-mean, circular symmetric complex random Gaussian vector that models $d$ Gaussian CV input states, $\mathcal{CN}(0,\mathbf{K_z})$, $\mathbf{K_z} = \mathbb{E}[\mathbf{z}\mathbf{z}^\dagger]$, where $z_i = x_i + \mathrm{i}p_i$, $\mathbf{x} = (x_0,\ldots,x_{d-1})^T$, $\mathbf{p} = (p_0,\ldots,p_{d-1})^T$, $x_i \in \mathbb{N}(0,\sigma^2_{\omega_0})$, $p_i \in \mathbb{N}(0,\sigma^2_{\omega_0})$ i.i.d. zero-mean Gaussian random variables. |
| $\mathbf{d} = F^{-1}(\mathbf{z})$ | An $l$-dimensional, zero-mean, circular symmetric complex random Gaussian vector, $\mathcal{CN}(0,\mathbf{K_d})$, $\mathbf{K_d} = \mathbb{E}[\mathbf{d}\mathbf{d}^\dagger]$, $\mathbf{d} = (d_0,\ldots,d_{l-1})^T$, $d_i = x_i + \mathrm{i}p_i$, $x_i, p_i \in \mathbb{N}(0,\sigma^2_{\omega_F})$ are i.i.d. zero-mean Gaussian random variables, $\sigma^2_{\omega_F} = 1/\sigma^2_{\omega_0}$. The $i$-th component is $d_i \in \mathcal{CN}(0,\sigma^2_{d_i})$, $\sigma^2_{d_i} = \mathbb{E}[|d_i|^2]$. |
| $\mathbf{y}_k \in \mathcal{CN}(0,\mathbb{E}[\mathbf{y}_k\mathbf{y}_k^\dagger])$ | A $d$-dimensional zero-mean, circular symmetric complex Gaussian random vector. |
| $y_{k,m}$ | The $m$-th element of the $k$-th user's vector $\mathbf{y}_k$, expressed as $y_{k,m} = \sum_l F(T_i(\mathcal{N}_i))F(d_i) + F(\Delta_i)$. |
| $F(\mathbf{T}(\mathcal{N}))$ | Fourier transform of $\mathbf{T}(\mathcal{N}) = [T_0(\mathcal{N}_0)\ldots T_{l-1}(\mathcal{N}_{l-1})]^T \in \mathcal{C}^l$, the complex |
| | transmittance vector. |
| $F(\Delta)$ | Complex vector, expressed as $F(\Delta) = e^{\frac{-F(\Delta)^T \mathbf{K}_{F(\Delta)} F(\Delta)}{2}}$, with covariance matrix $\mathbf{K}_{F(\Delta)} = \mathbb{E}\left[F(\Delta) F(\Delta)^\dagger\right]$. |
| $\mathbf{y}[j]$ | AMQD block, $\mathbf{y}[j] = F(\mathbf{T}(\mathcal{N})) F(\mathbf{d})[j] + F(\Delta)[j]$. |
| $\tau = \|F(\mathbf{d})[j]\|^2$ | An exponentially distributed variable, with density $f(\tau) = \left(1/2\sigma_\omega^{2n}\right) e^{-\tau/2\sigma_\omega^2}$, $\mathbb{E}[\tau] \leq n 2\sigma_\omega^2$. |
| $T_{Eve,i}$ | Eve's transmittance on the Gaussian sub-channel $\mathcal{N}_i$, $T_{Eve,i} = \operatorname{Re} T_{Eve,i} + \mathrm{i}\operatorname{Im} T_{Eve,i} \in \mathcal{C}$, $0 \leq \operatorname{Re} T_{Eve,i} \leq 1/\sqrt{2}$, $0 \leq \operatorname{Im} T_{Eve,i} \leq 1/\sqrt{2}$, $0 \leq |T_{Eve,i}|^2 < 1$. |
| $d_i$ | A $d_i$ subcarrier in an AMQD block. |
| $\nu_{\min}$ | The $\min\{\nu_0,\ldots,\nu_{l-1}\}$ minimum of the $\nu_i$ sub-channel coefficients, where $\nu_i = \sigma_\mathcal{N}^2 \big/ |F(T_i(\mathcal{N}_i))|^2$ and $\nu_i < \nu_{Eve}$. |
| $\sigma_\omega^2$ | Constant modulation variance, $\sigma_\omega^2 = \nu_{Eve} - \nu_{\min} \mathcal{G}(\delta)_{p(x)}$, $\nu_{Eve} = \frac{1}{\lambda}$, $\lambda = |F(T_\mathcal{N}^*)|^2 = \frac{1}{n} \sum_{i=0}^{n-1} \left|\sum_{k=0}^{n-1} T_k^* e^{\frac{-\mathrm{i}2\pi i k}{n}}\right|^2$ and $T_\mathcal{N}^*$ is the expected transmittance of the Gaussian sub-channels under an optimal Gaussian collective attack. |

## S.2 Abbreviations

| | |
|---|---|
| AMQD | Adaptive Multicarrier Quadrature Division |
| AWGN | Additive White Gaussian Noise |
| CV | Continuous-Variable |
| CVQFT | Continuous-Variable Quantum Fourier Transform |
| CVQKD | Continuous-Variable Quantum Key Distribution |
| DGQI | Direct Gaussian Quadrature Inference |
| DV | Discrete Variable |
| FFT | Fast Fourier Transform |
| GQI | Gaussian Quadrature Inference |
| ICVQFT | Inverse Continuous-Variable Quantum Fourier Transform |
| IFFT | Inverse Fast Fourier Transform |
| MQA | Multiuser Quadrature Allocation |
| PDF | Probability Density Function |
| QKD | Quantum Key Distribution |
| SNR | Signal to Noise Ratio |
| WSS | Wide-Sense Stationary |